\documentclass{elsart}

\usepackage{graphicx}
\usepackage{epsfig}

\usepackage{amssymb}
\usepackage{rotating}

\begin{document}
\begin{frontmatter}
\title {Evaluation of $(\alpha, n)$ Induced Neutrons as a Background for Dark Matter Experiments }
\author[usd]{D.-M. Mei\corauthref{cor}},
\corauth[cor]{Corresponding author.}
\ead{dongming.mei@usd.edu}
\author[usd,ctgu]{C. Zhang},
\author[lanl]{A. Hime}
\address[usd]{Department of Physics, The University of South Dakota, 
Vermillion, South Dakota 57069}
\address[ctgu]{College of Sciences, China Three Gorges University, Yichang 443002, China}
\address[lanl]{P-23, H803, Los Alamos National Laboratory, Los Alamos, New Mexico 87545}
\begin{abstract}
% Text of abstract
Neutrons from ($\alpha$,n) reactions through thorium and uranium decays are important sources of  background for direct
dark matter detection.  The neutron yields and energy spectra from a range of materials that are used to build 
dark matter detectors are calculated and tabulated.  In addition to thorium and uranium decays, we found
that $\alpha$ particles from samarium, often the dopant of the window materials of photomultiplier tubes (PMT), are also an 
important source of neutron yield. The results in this paper can be used as the input to Monte Carlo simulations 
for many materials that will be used for next generation experiments. 

\end{abstract}

\begin{keyword}
$(\alpha,n)$ neutrons \sep Dark matter detection 

\PACS 13.85.Tp \sep 23.40-s \sep 25.40.Sc \sep 28.41.Qb \sep 95.35.+d \sep 29.40.Wk
\end{keyword}
\end{frontmatter}

\section{Introduction}
Neutron induced elastic scattering processes represent an important background 
for direct dark matter detection experiments 
searching for Weakly Interacting Massive Particles (WIMPs), 
which may constitute the dark matter in the 
universe~\cite{dns, wfr, mwg, gjm}. Direct searches for WIMPs
have been carried out by many experiments including CDMS~\cite{cdms1}, EDELWEISS~\cite{edel1}, Xenon10~\cite{xenon10}, 
ArDM~\cite{argon}, DAMA~\cite{dama1}, CRESST~\cite{cresst1}, PICASSO~\cite{pica1}, NAIAD~\cite{naia}, and ZEPLIN~\cite{zep1}.
Among these experiments, DAMA/NaI~\cite{bern1}  and DAMA/LIBRA~\cite{bern2} have claimed that they have observed 
a model independent annual modulation signature. The DAMA collaboration interprets 
this annual modulation as the signature induced by dark matter (DM) 
particles~\cite{bern1,bern2}.  However, this claim is at odds with other experimental results if one assumes standard WIMP interactions and halo
models. 
With the best limits set by CDMS II~\cite{cdms2}
and Xenon10, WIMPs remain unobserved. Large scale next generation detectors utilizing noble liquids to continue the direct search for 
WIMPs are underway. 
The key to these experiments lies in the ability to reduce various
background to unprecedented low levels.
Among all possible sources of background, neutron induced nuclear recoil is identified as a major
 source of 
background for this type of experiments. 

There are three sources of underground neutrons: 1) those produced by ($\alpha,n)$ reactions through 
thorium, uranium, and other radioactive isotope decays in the materials that surround or constitute the detector; 2)
those from spontaneous uranium fission; and 3) those from cosmic ray muon-induced processes. In general, the ($\alpha,n)$
neutrons dominate the total neutron contributions to the measured background for an underground experiment.
This is because the origins of ($\alpha, n)$ neutrons range from surrounding rock, external shielding, 
inner shielding, detector components, and detector target. In particular, the ($\alpha,n)$ neutrons produced
in the detector components and target are hard to cope with. These neutrons need to be understood very well
in terms of their origin, transport, and interaction with materials. Calculations of the
neutron yield and neutron energy spectrum in different materials are critical to dark matter experiments. 
The total neutron yield indicates the number of neutrons that enter or are produced in the target. The neutron energy spectrum
determines the total background events in the region of interest (ROI).  Therefore, a full simulation of 
neutron background must take into account both neutron yield and energy spectrum. This paper is aimed at
providing the ($\alpha,n)$ neutron yield and energy spectrum for a number of materials that are used for
the construction of dark matter detectors.

\section{Calculations of neutron yields and energy spectra}
The neutron yields from the ($\alpha,n)$ reaction for various elements of natural isotopic concentrations have been discussed 
by many authors~\cite{wes,bai,lis, fei, heat1990, heat, sha, cjh, jhg, wfi, lva, kev, erb, rou}. 
The decays of $^{238}U$ and $^{232}Th$ in the materials produce MeV $\alpha$ particles.
These $\alpha$ particles interact with the nucleus
in a thick target and yield neutrons. The neutron yield is calculated by~\cite{heat1990}:
\begin{equation}\label{yield}
Y_{i} = \frac{N_{A}}{A_{i}}\int^{E_{0}}_{0}\frac{\sigma_{i}(E)}{S^{m}_{i}(E)}\rm{d}E,
\end{equation}
where $E_{0}$ is the initial energy of the $\alpha$-particle, $S^{m}_{i}$ is the mass stopping power of element 
$i$, $A_{i}$ is the atomic mass of element $i$ and $N_{A}$ is Avogadro's constant.
In secular equilibrium, the $^{232}$Th decay chain yields 6 $\alpha$'s and the $^{238}$U decay chain produces 
8 $\alpha$'s with various energies $E_{j}$. 
The neutron yields in the decay chains of $^{232}$Th and $^{238}$U can be determined by the sum of the individual yields
induced by 
each $\alpha$, weighted by the branching ratio for each element and weighted by the mass ratio 
in the host material. The energy attenuation of the $\alpha$-particles in the medium is the dominant process
for the thick target hypothesis. 
Under the assumption that the incident flux of $\alpha$-particles with energy of $E_{j}$
  is invariant until the energy is attenuated to zero,  
  for target element $i$,
 the differential spectra of neutron yield can be
expressed as
\begin{eqnarray}
Y_{i}(E_{n}) &= &N_{i}\sum_{j}\Phi_{\alpha}(E_{j})
  \int_{0}^{E_{j}}\frac{\rm{d}\sigma(E_{\alpha}, E_{n})}{\rm{d}E_{\alpha}}\rm{d}E_{\alpha} \nonumber \\
   &= &\frac{N_{A}}{A_{i}}\sum_{j} \frac{R_{\alpha}(E_{j})}{S_{i}^{m}(E_{j})}
  \int_{0}^{E_{j}}\frac{\rm{d}\sigma(E_{\alpha}, E_{n})}{\rm{d}E_{\alpha}}\rm{d}E_{\alpha}, \label{diff} 
\end{eqnarray}
where $N_{i}$ is the total number of atoms for the i$^{th}$ element in the host material,
$\Phi_{\alpha}(E_{j})$ is the flux of $\alpha$-particles with specific energy $E_{j}$,
 $R_{\alpha}(E_{j})$ refers to the $\alpha$-particle production rate
 for the decay with the energy $E_{j}$ 
from $^{232}$Th or $^{238}$U decay chain. If we consider the specific activity of $^{232}$Th and $^{238}$U 
in terms of the concentration in ppm/g/y, then 
\begin{equation}
R_{\alpha}(E_{j}) = 10^{-6}\frac{N_{A}}{A_{a}}\frac{\ln2}{t_{1/2}}B_{j},
\label{added}
\end{equation}
where $A_{a}$ stands for the atomic mass number of $^{232}$Th or $^{238}$U,
 $B_{j}$ represents the $\alpha$-particle 
branching ratio for a specific energy decay channel $E_{j}$, and $ t_{1/2}$ is the half life of the decay.
\par
The cross section in Eq.(\ref{diff}),
 $\int_{0}^{E_{j}}\frac{\rm{d}\sigma(E_{\alpha}, E_{n})}{\rm{d}E_{\alpha}}\rm{d}E_{\alpha} $,
 is calculated by the TALYS simulation code~\cite{talys}  in which the cross sections of neutron production 
for all possible reaction channels are calculated. 
The flux of $\alpha$-particles is obtained by combining 
the production rate of $\alpha$-particles with the corresponding mass stopping power in the 
target. The mass stopping power for specific energies is calculated by our simulation described in Ref.~\cite{mei} and 
the ASTAR program~\cite{astar}. 
We use the decay chains of selected isotopes in Ref.~\cite{branch} for 
$\alpha$-particle emission from $^{238}U$ and $^{232}Th$ decays. Only decays with visible energies larger than
0.1 MeV or branching ratio more than 0.5\% are included.    

\section{Results and Discussions}
\par
\begin{figure}
\centering
\includegraphics[width=0.9\textwidth]{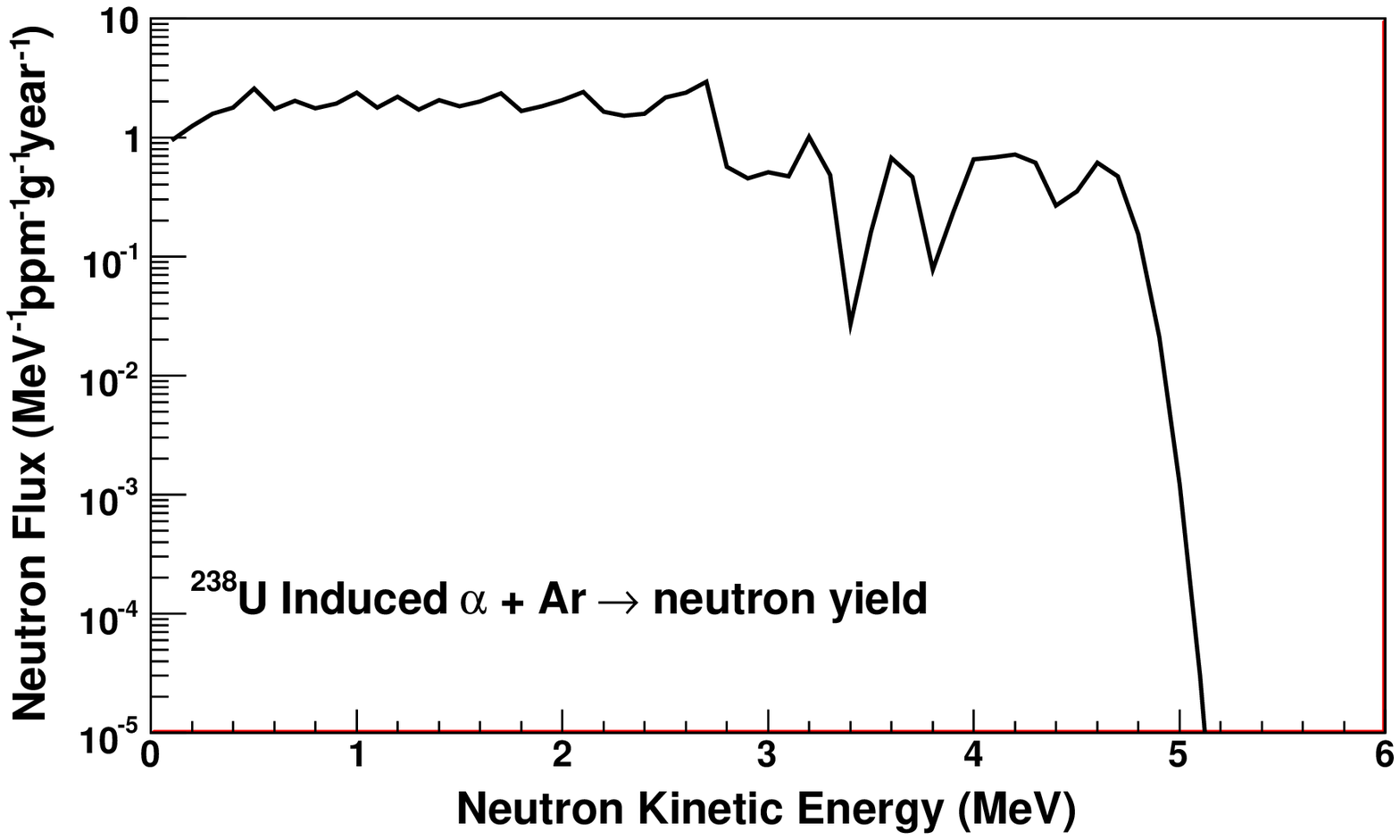}
\includegraphics[width=0.9\textwidth]{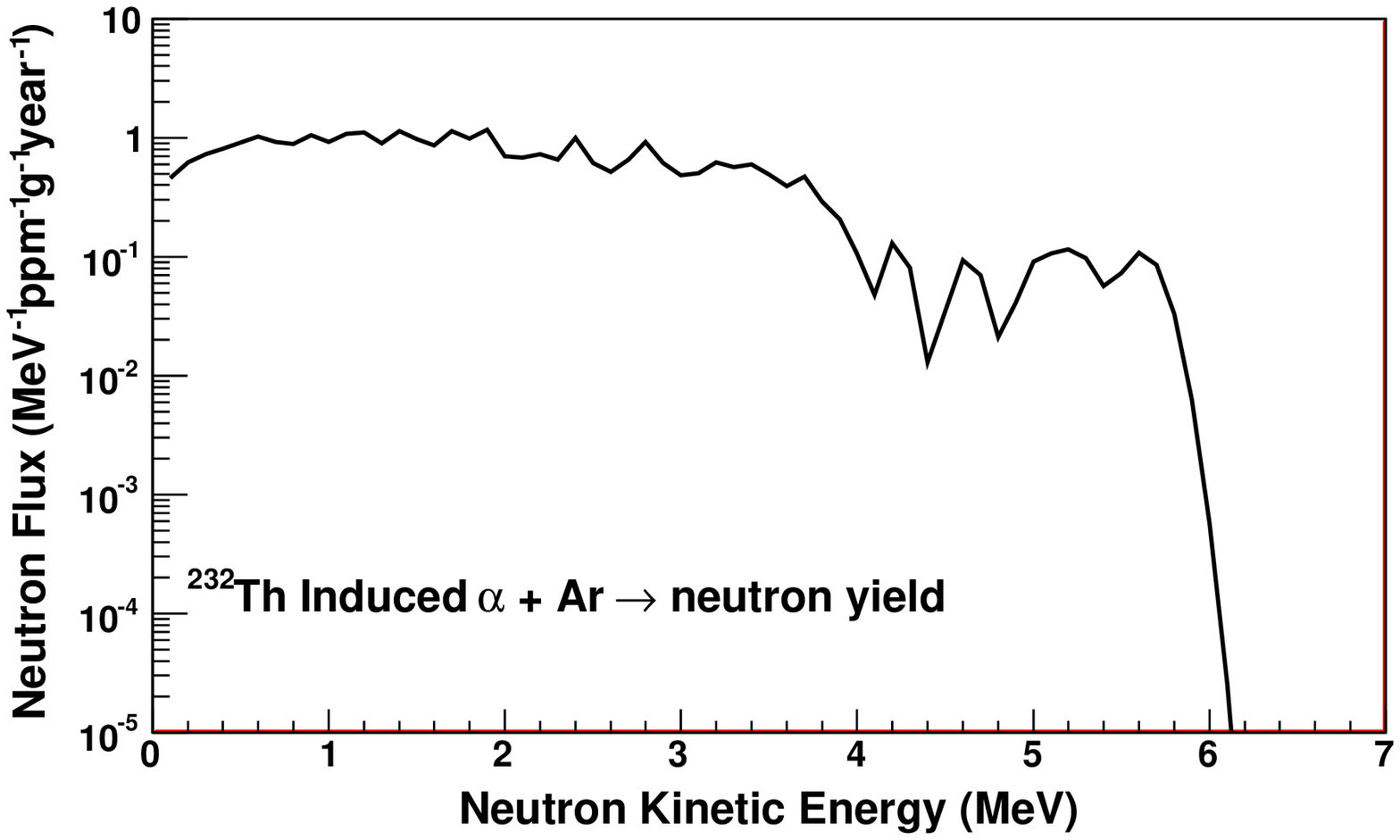}
\caption{The differential neutron flux induced by $(\alpha, n)$ reaction in a thick target of argon. The $\alpha$-particles
  are induced by $^{238}U$ and $^{232}Th$ decays.}\label{fig:Ar}
\end{figure}
\begin{figure}
\centering
\includegraphics[width=0.9\textwidth]{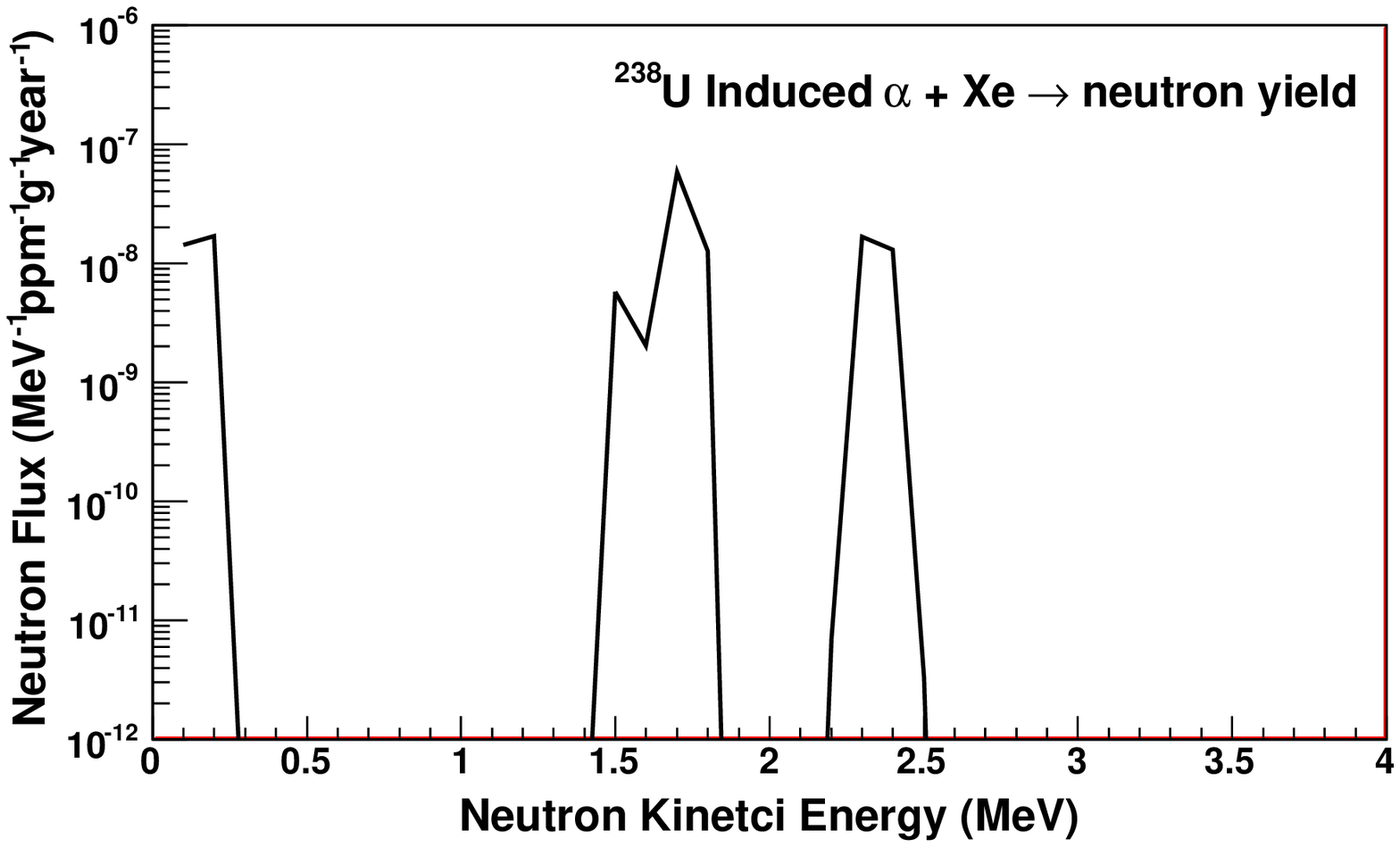}
\includegraphics[width=0.9\textwidth]{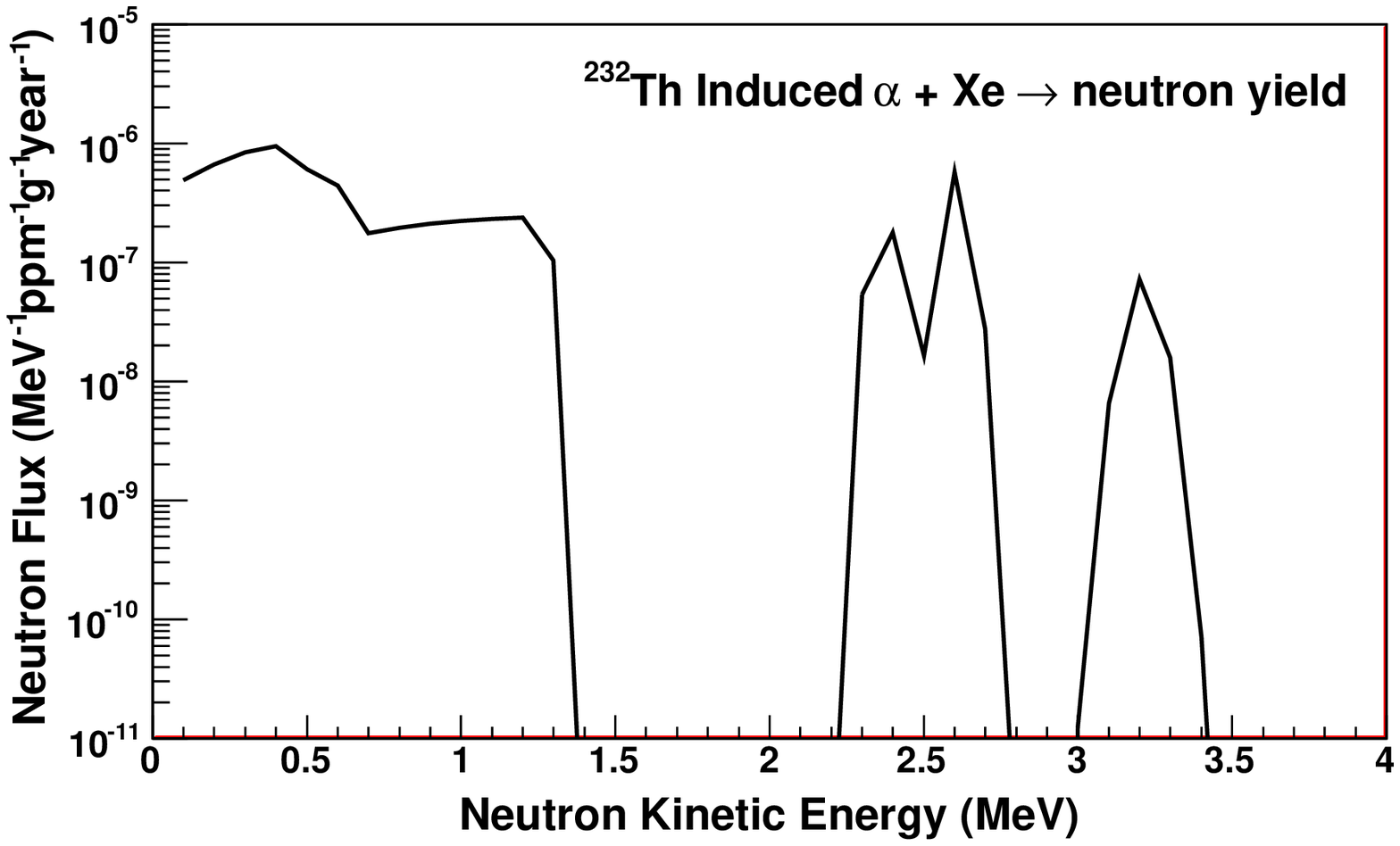}
\caption{The differential neutron flux induced by $(\alpha, n)$ reaction in a thick target of xenon. The $\alpha$-particles
  are induced by $^{238}U$ and $^{232}Th$ decays.}\label{fig:Xe}
\end{figure}
\begin{figure}
\centering
\includegraphics[width=0.9\textwidth]{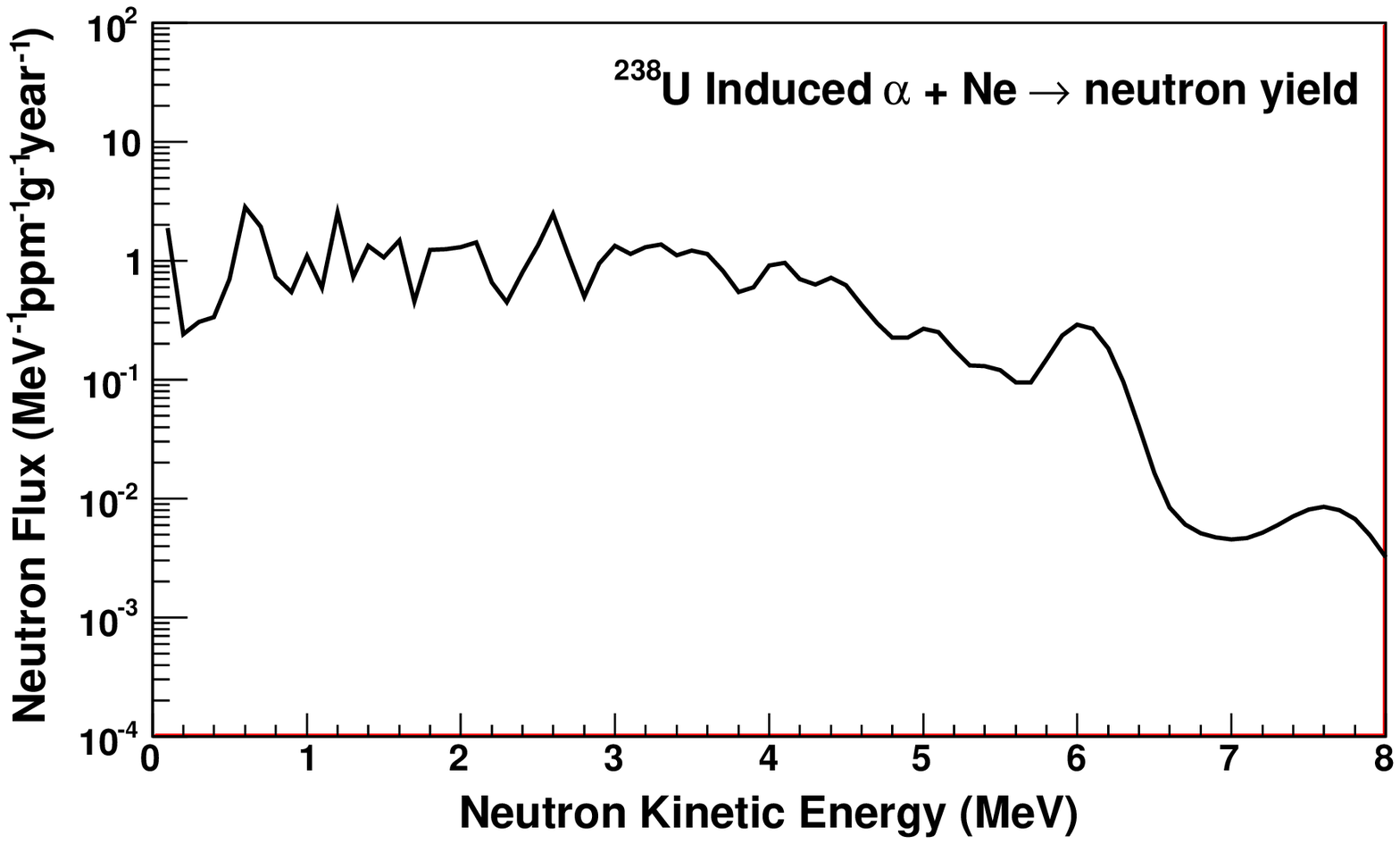}
\includegraphics[width=0.9\textwidth]{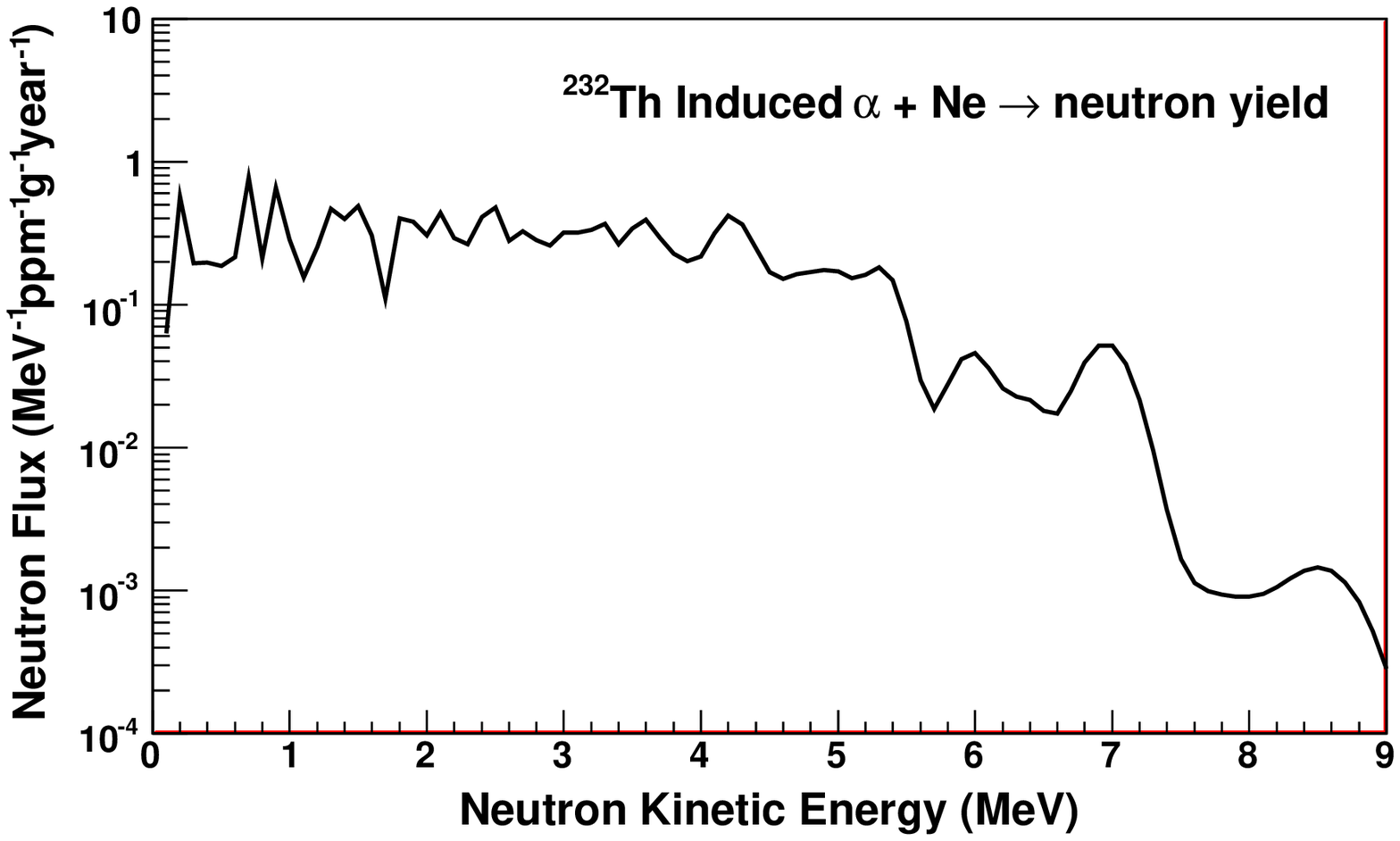}
\caption{The differential neutron flux induced by $(\alpha, n)$ reaction in a thick target of neon. The $\alpha$-particles
  are induced by $^{238}U$ and $^{232}Th$ decays.}\label{fig:Ne}
\end{figure}
\begin{figure}
\centering
\includegraphics[width=0.9\textwidth]{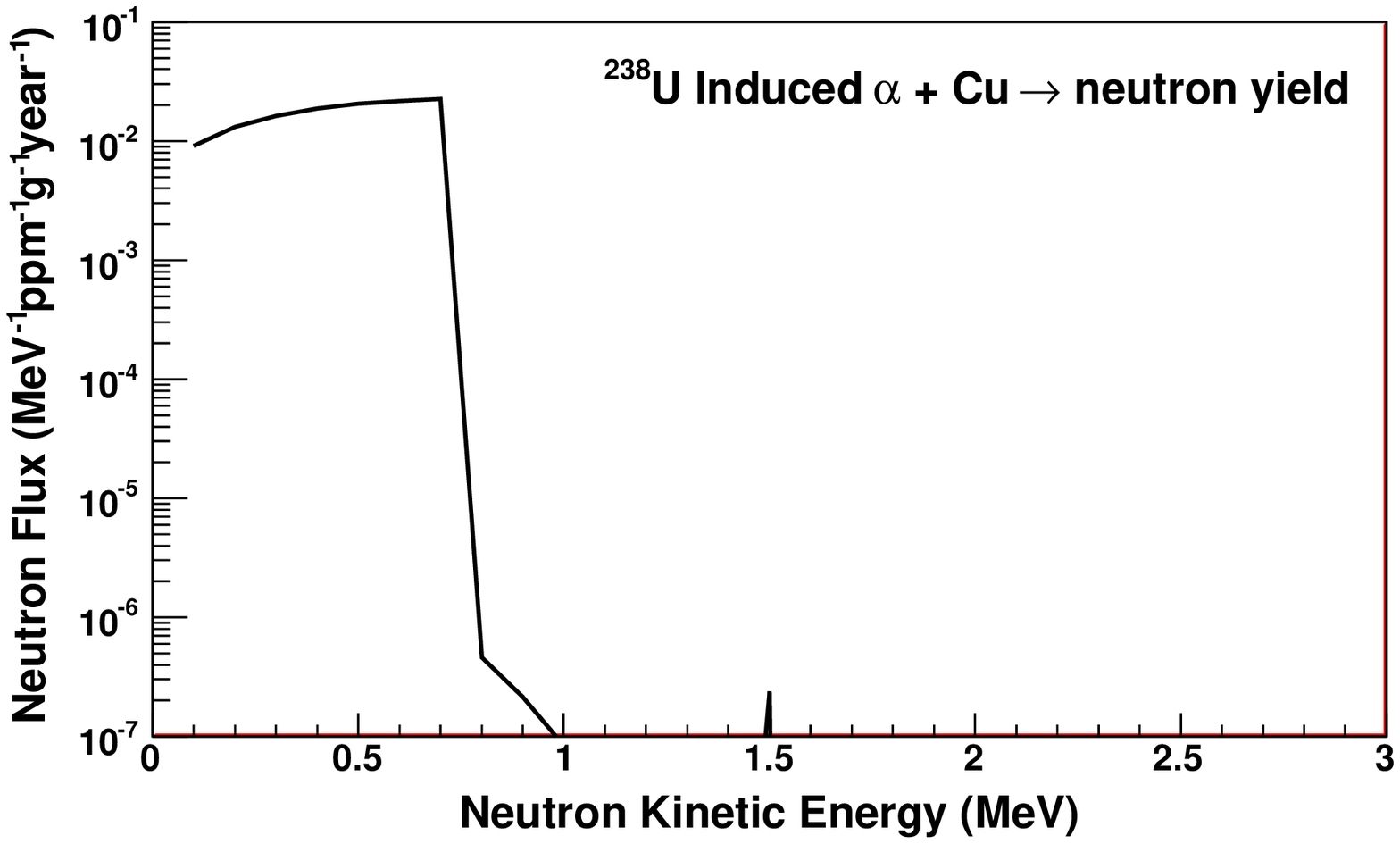}
\includegraphics[width=0.9\textwidth]{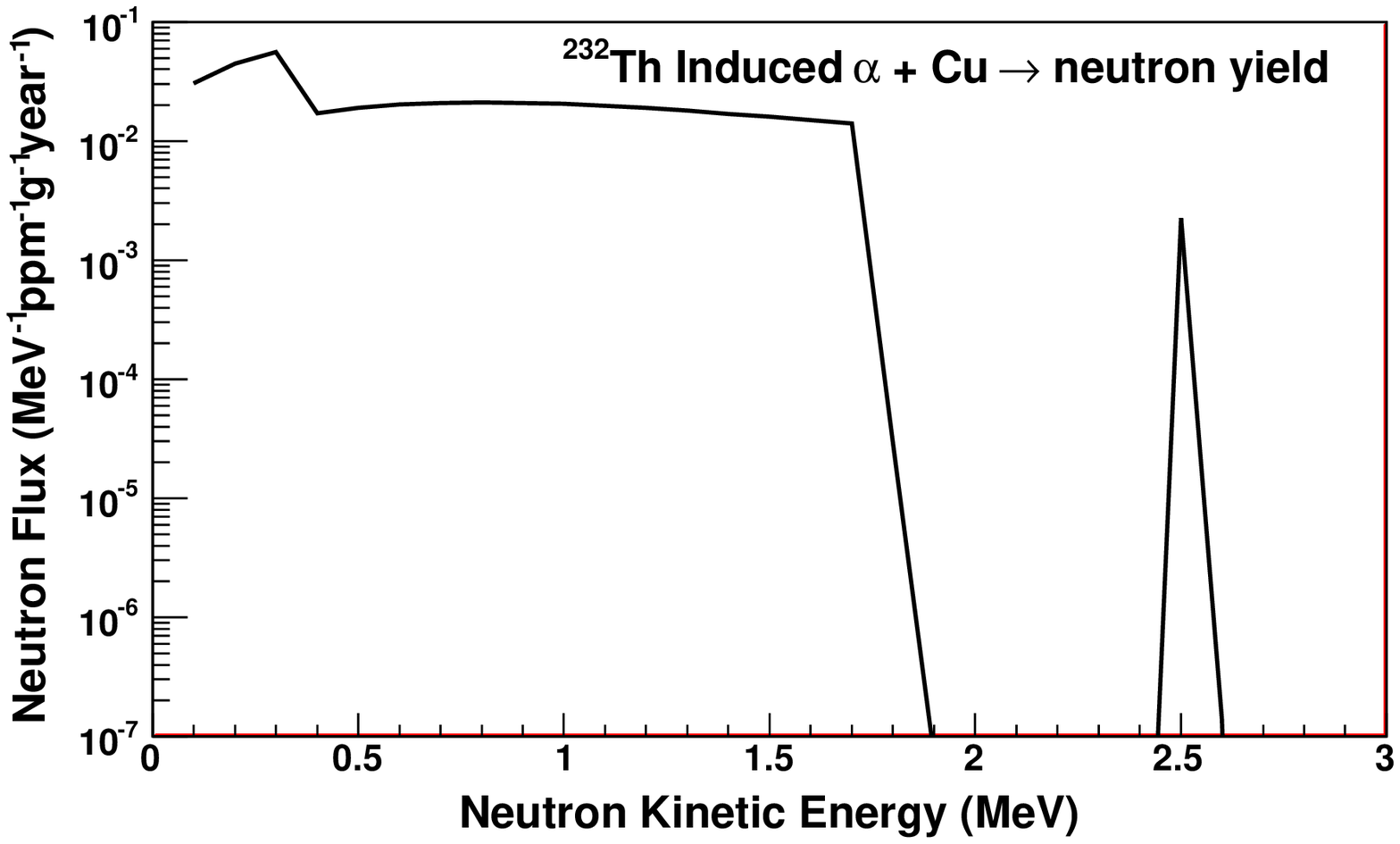}
\caption{The differential neutron flux induced by $(\alpha, n)$ reaction in a thick target of copper. The $\alpha$-particles
  are induced by $^{238}U$ and $^{232}Th$ decays.}\label{fig:Cu}
\end{figure}
\begin{figure}
\centering
\includegraphics[width=0.9\textwidth]{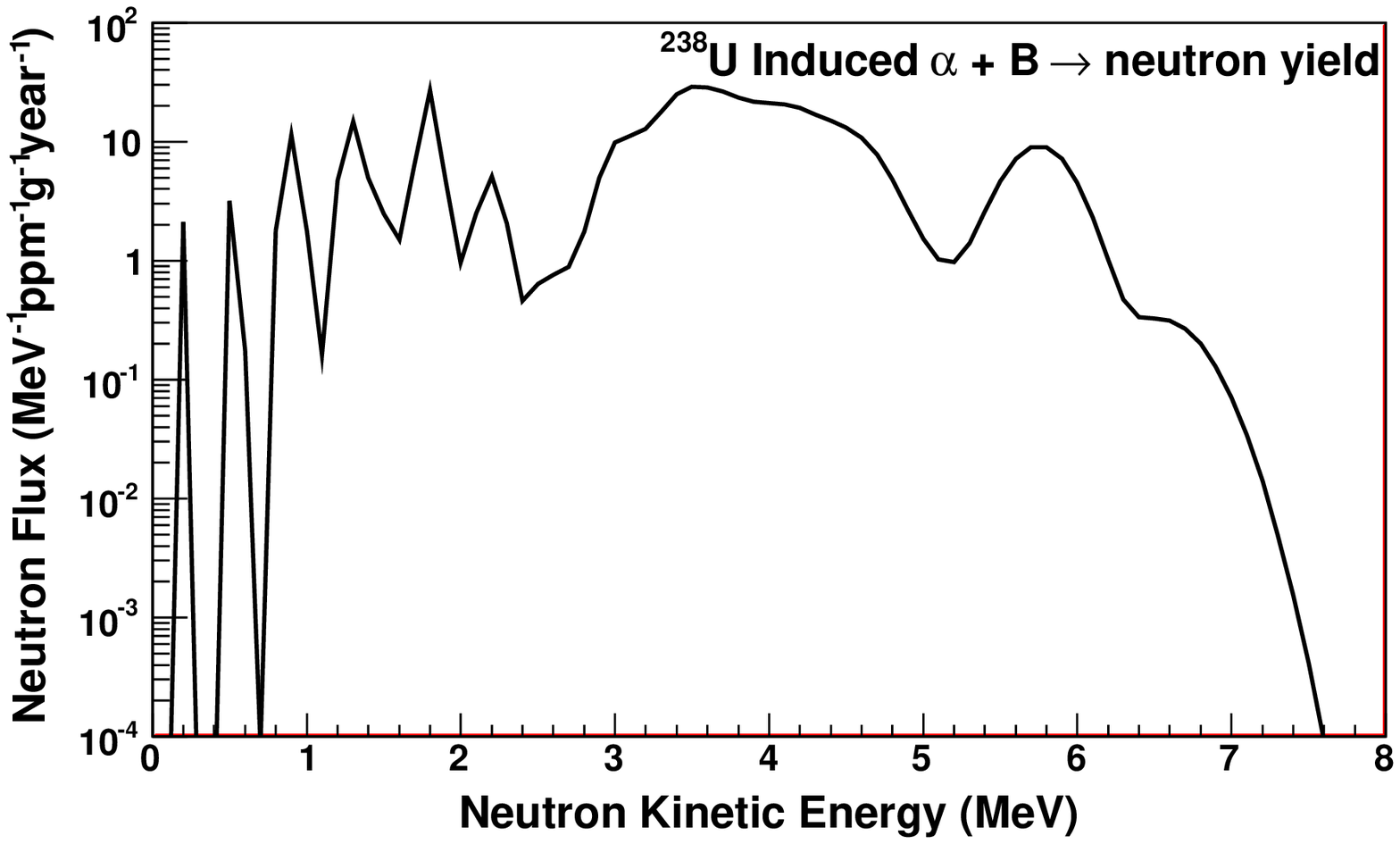}
\includegraphics[width=0.9\textwidth]{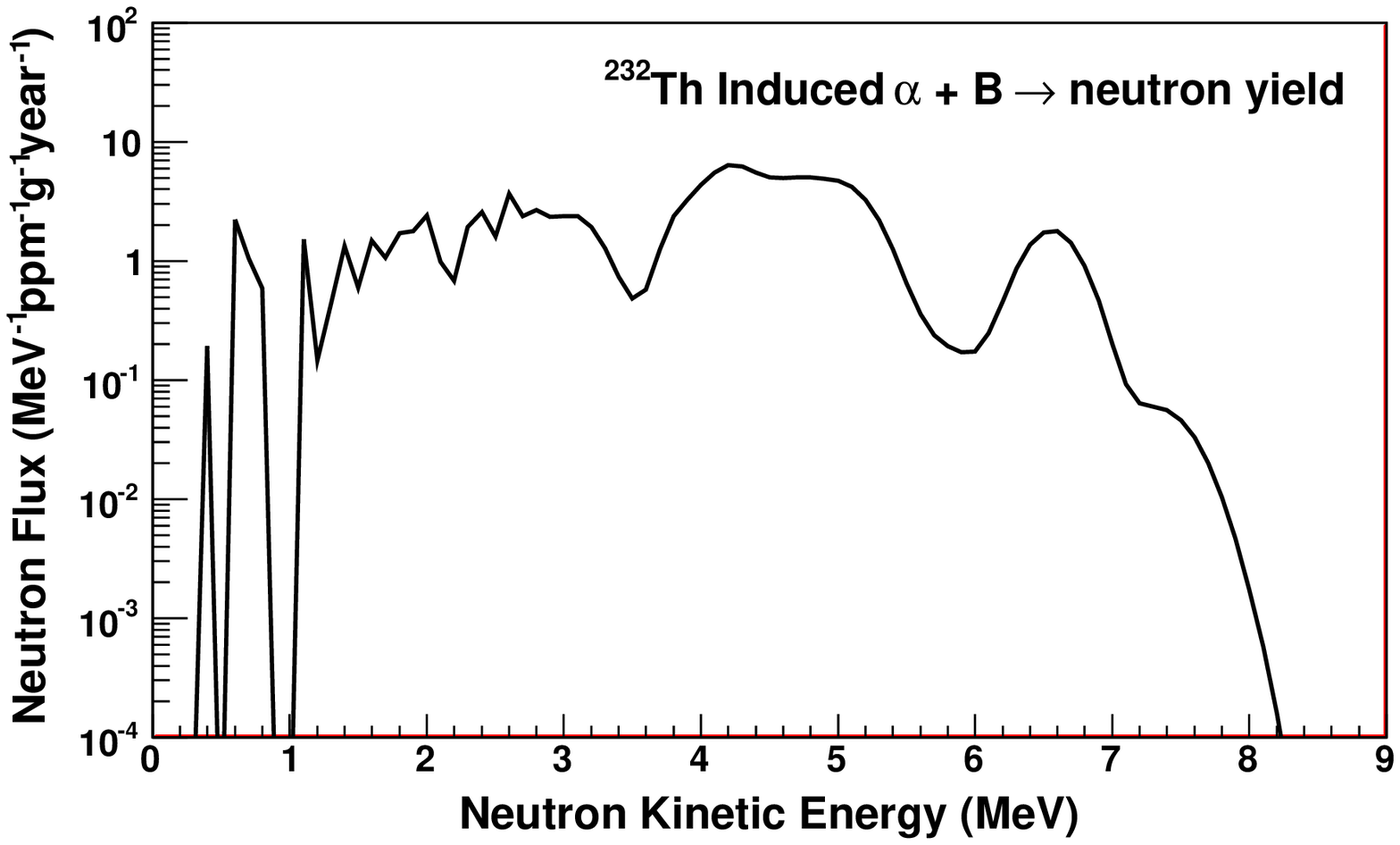}
\caption{The differential neutron flux induced by $(\alpha, n)$ reaction in a thick target of boron. The $\alpha$-particles
  are induced by $^{238}U$ and $^{232}Th$ decays.}\label{fig:B}
\end{figure}
\begin{figure}
\centering
\includegraphics[width=0.9\textwidth]{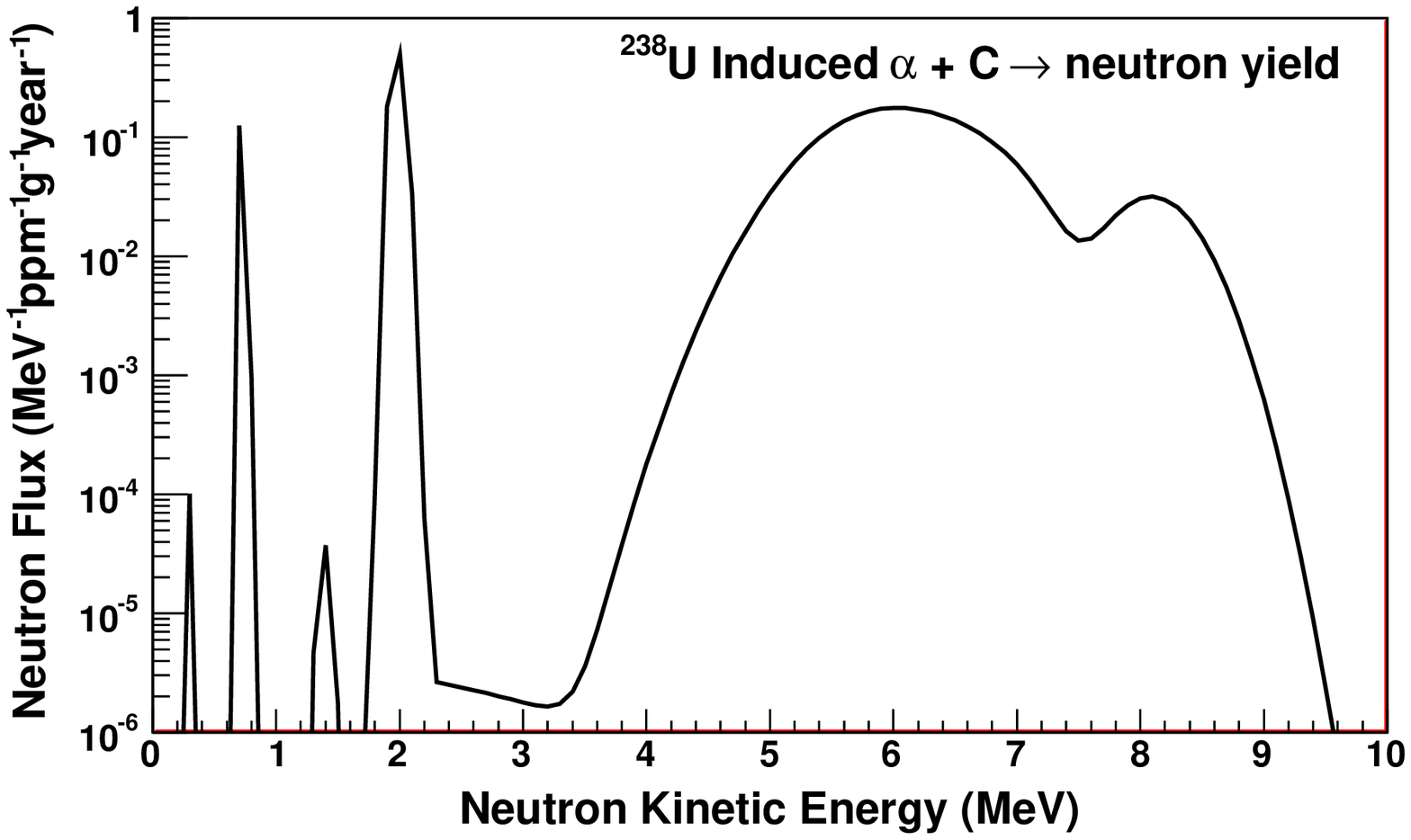}
\includegraphics[width=0.9\textwidth]{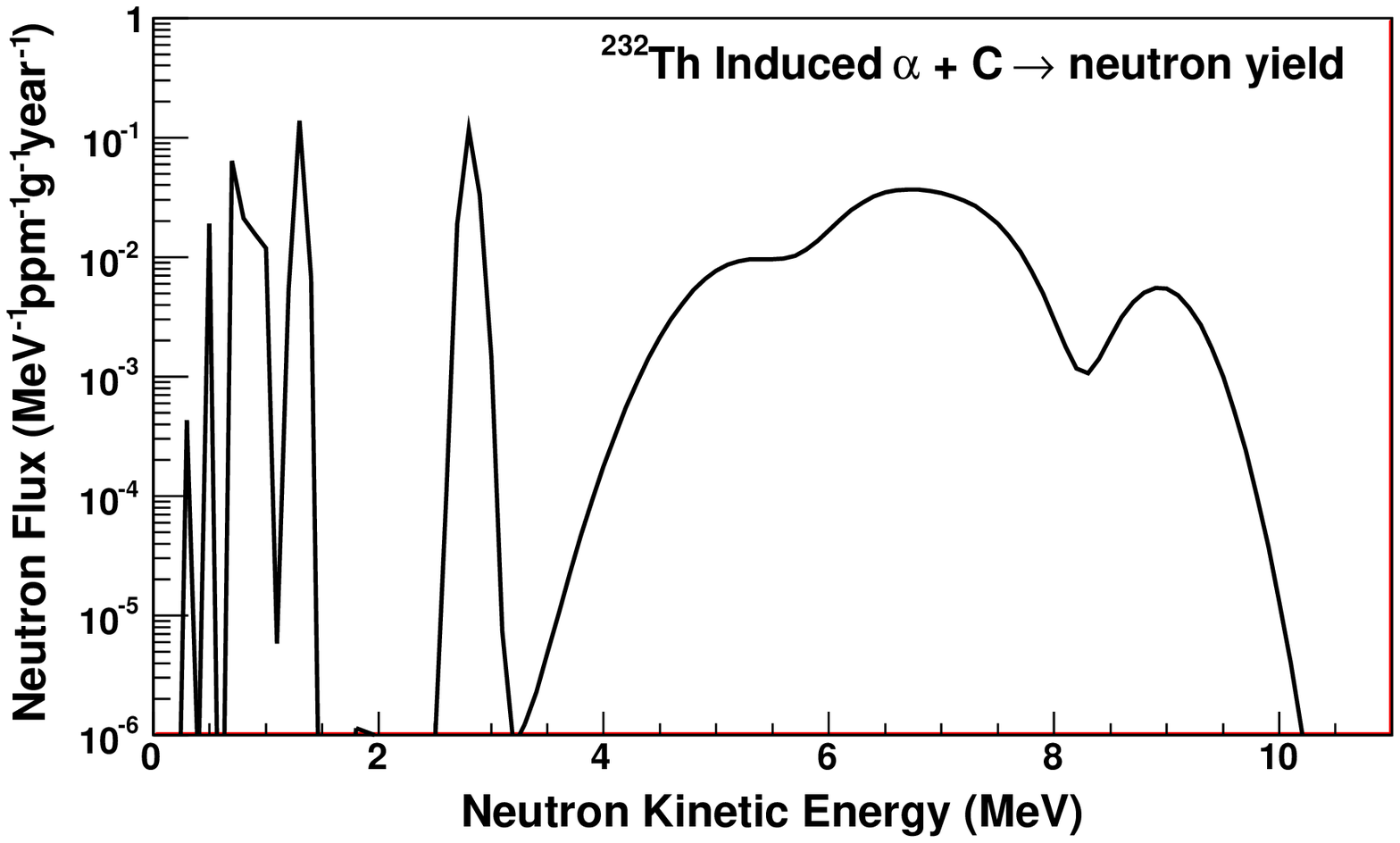}
\caption{The differential neutron flux induced by $(\alpha, n)$ reaction in a thick target of carbon. The $\alpha$-particles
  are induced by $^{238}U$ and $^{232}Th$ decays.}\label{fig:C}
\end{figure}
\begin{figure}
\centering
\includegraphics[width=0.9\textwidth]{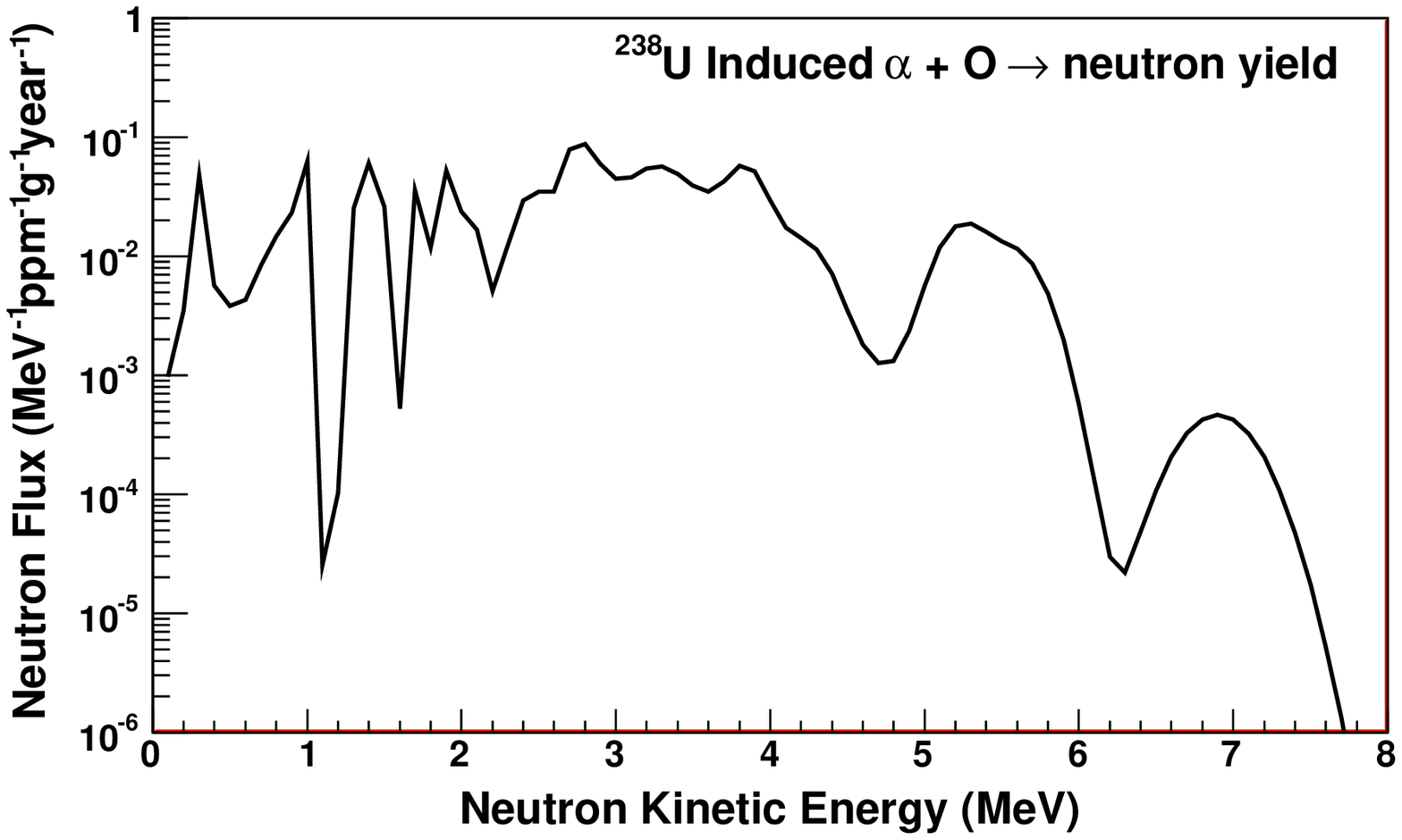}
\includegraphics[width=0.9\textwidth]{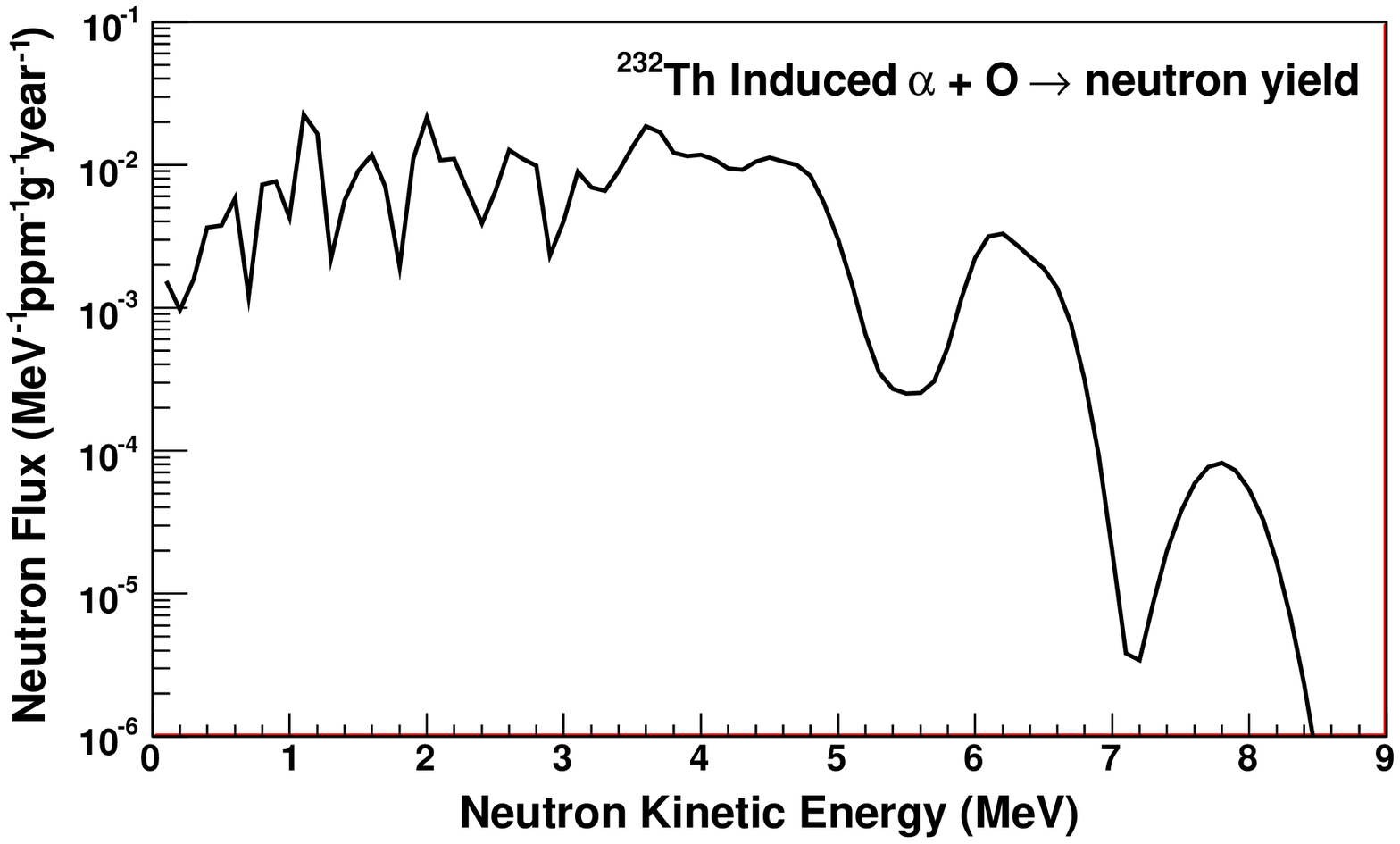}
\caption{The differential neutron flux induced by $(\alpha, n)$ reaction in a thick target of oxygen. The $\alpha$-particles
  are induced by $^{238}U$ and $^{232}Th$ decays.}\label{fig:O}
\end{figure}
\begin{figure}
\centering
\includegraphics[width=0.9\textwidth]{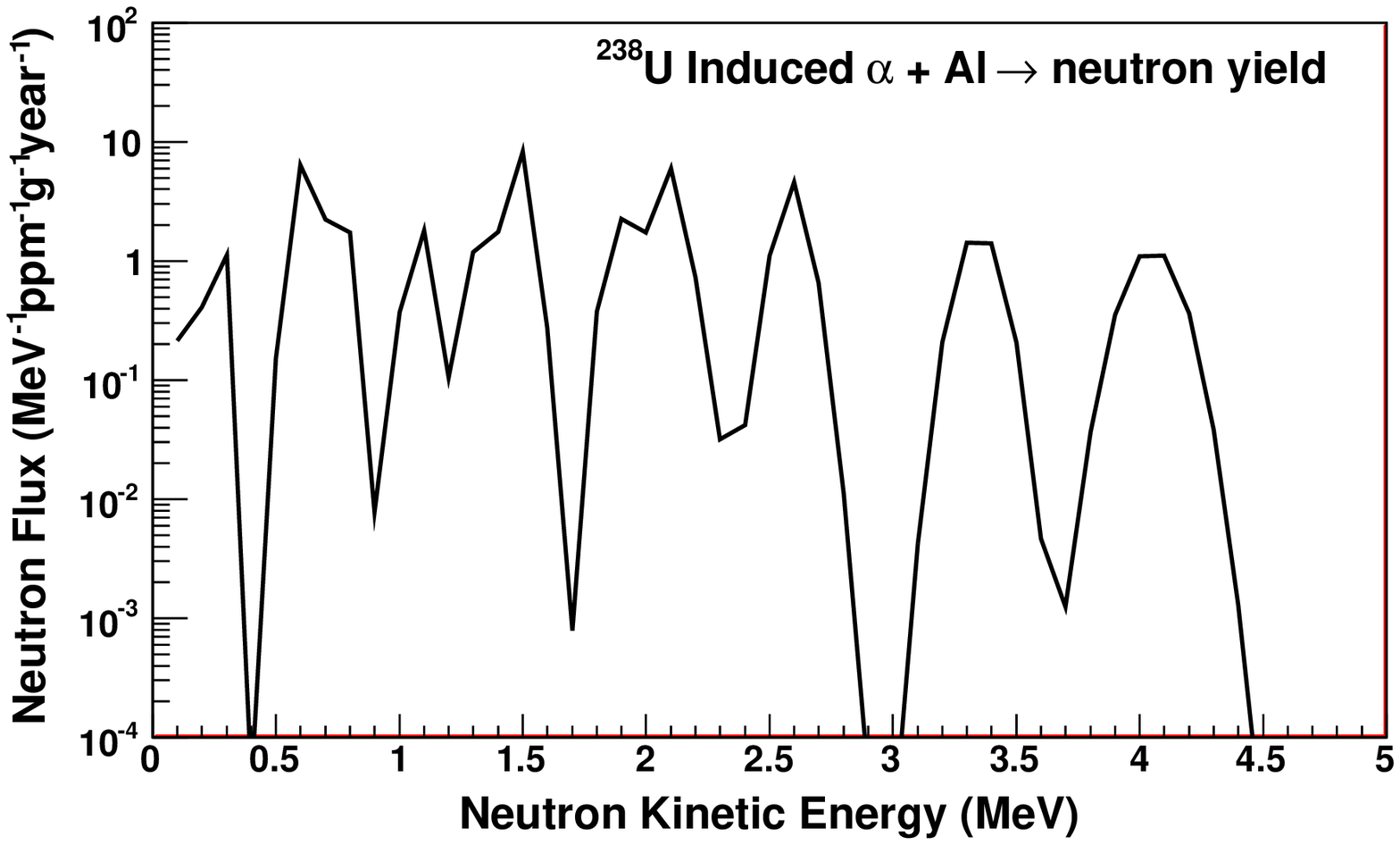}
\includegraphics[width=0.9\textwidth]{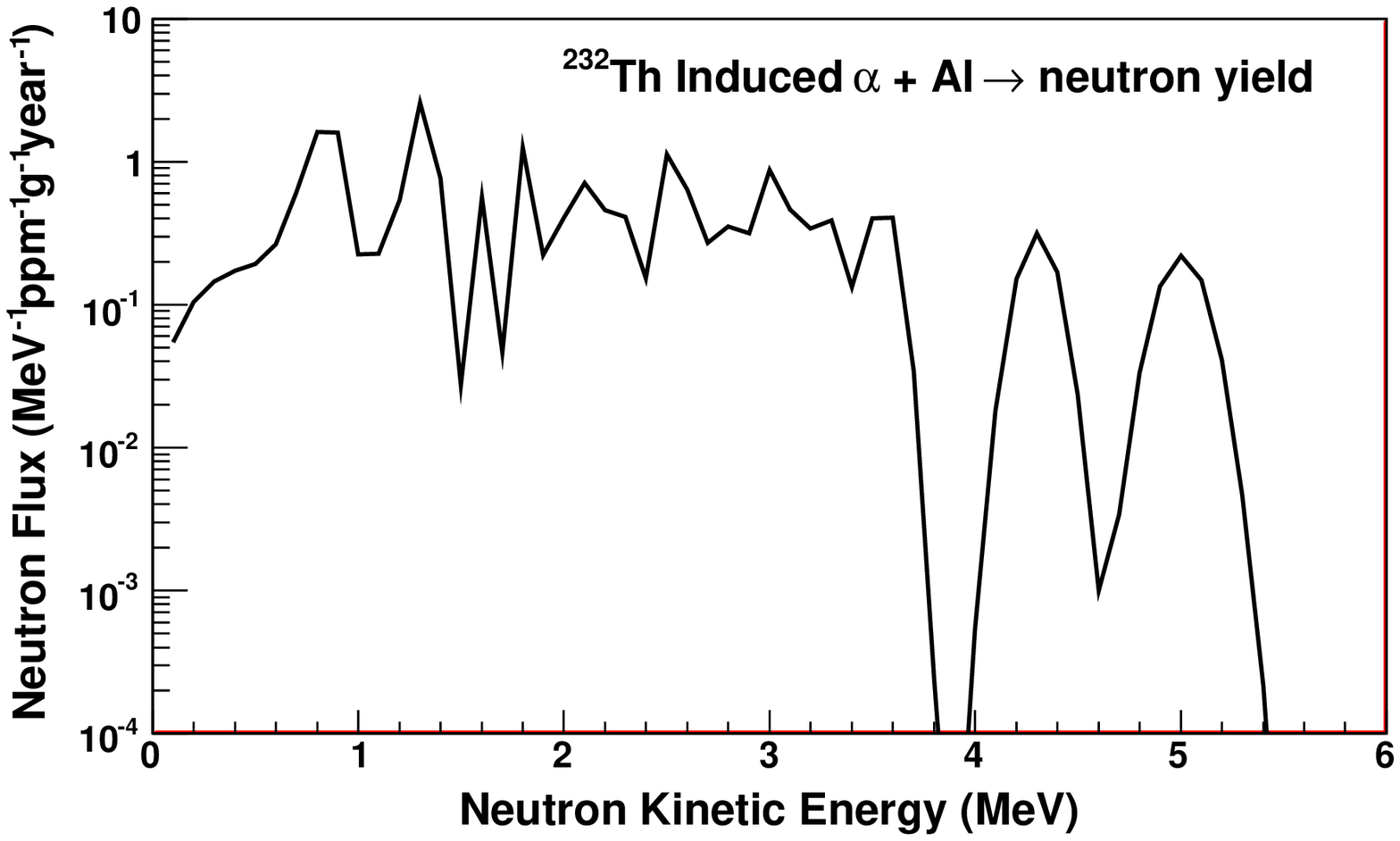}
\caption{The differential neutron flux induced by $(\alpha, n)$ reaction in a thick target of aluminum. The $\alpha$-particles
  are induced by $^{238}U$ and $^{232}Th$ decays.}\label{fig:Al}
\end{figure}
\begin{figure}
\centering
\includegraphics[width=0.9\textwidth]{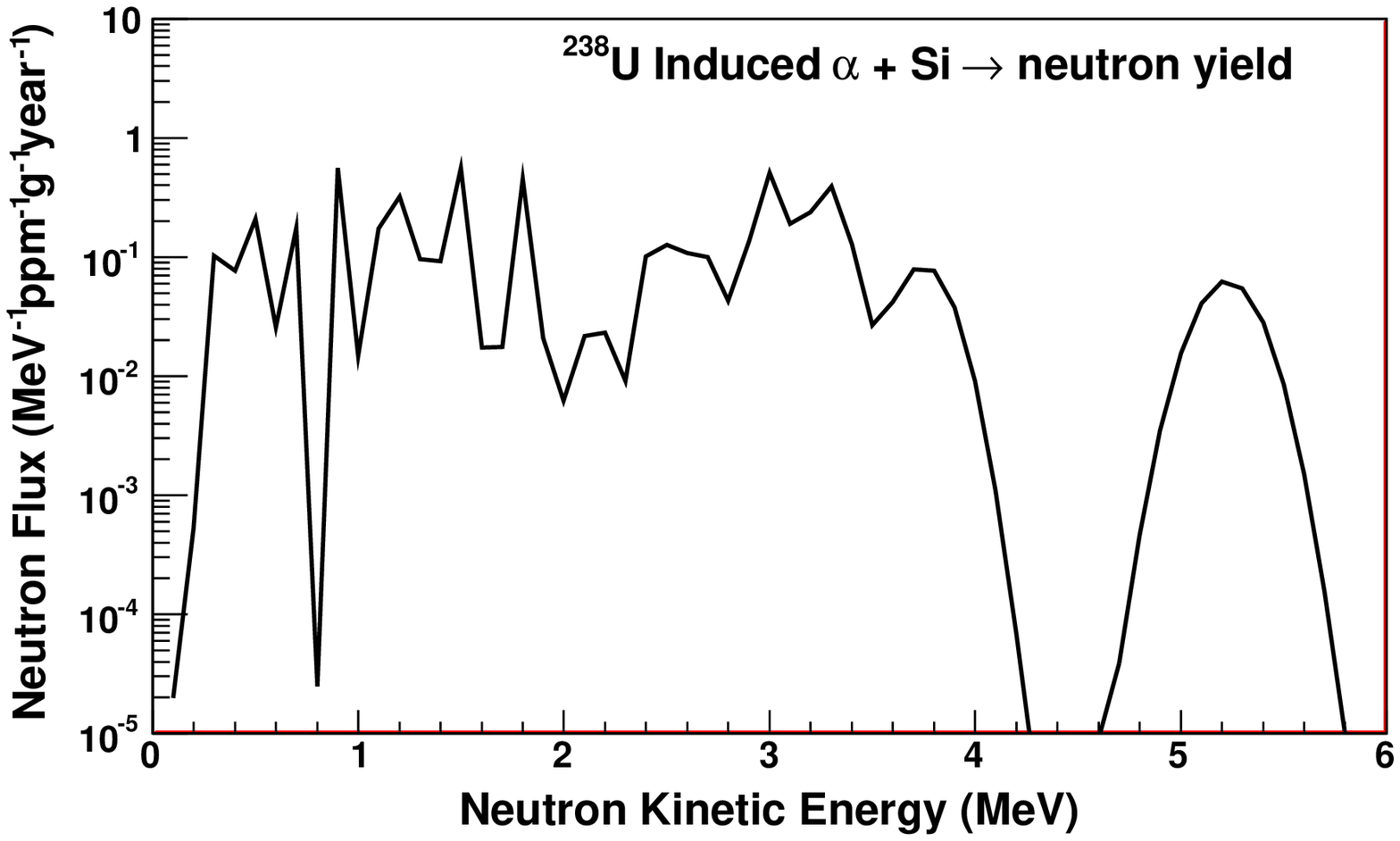}
\includegraphics[width=0.9\textwidth]{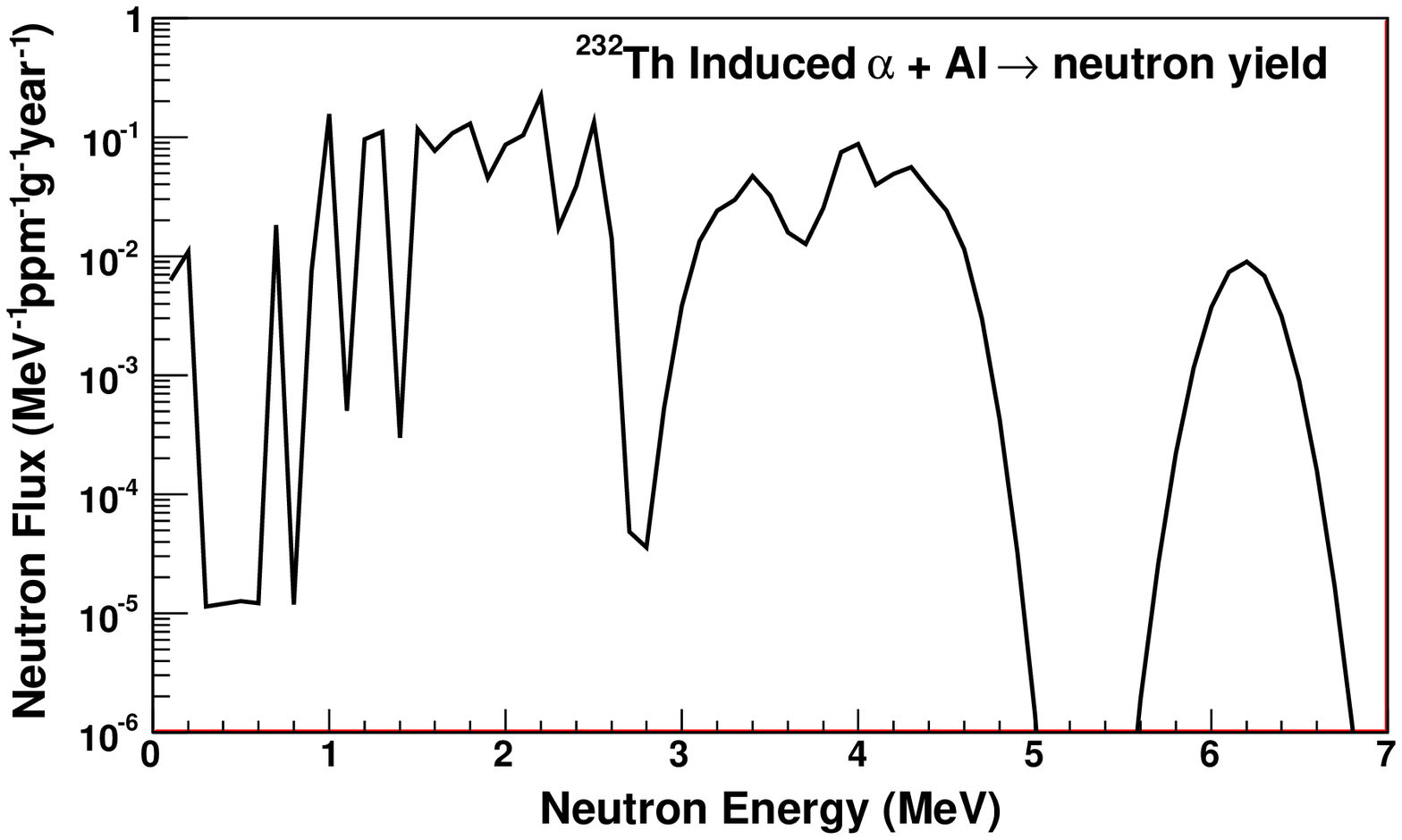}
\caption{The differential neutron flux induced by $(\alpha, n)$ reaction in a thick target of silicon. The $\alpha$-particles
  are induced by $^{238}U$ and $^{232}Th$ decays.}\label{fig:Si}
\end{figure}
\begin{figure}
\centering
\includegraphics[width=0.9\textwidth]{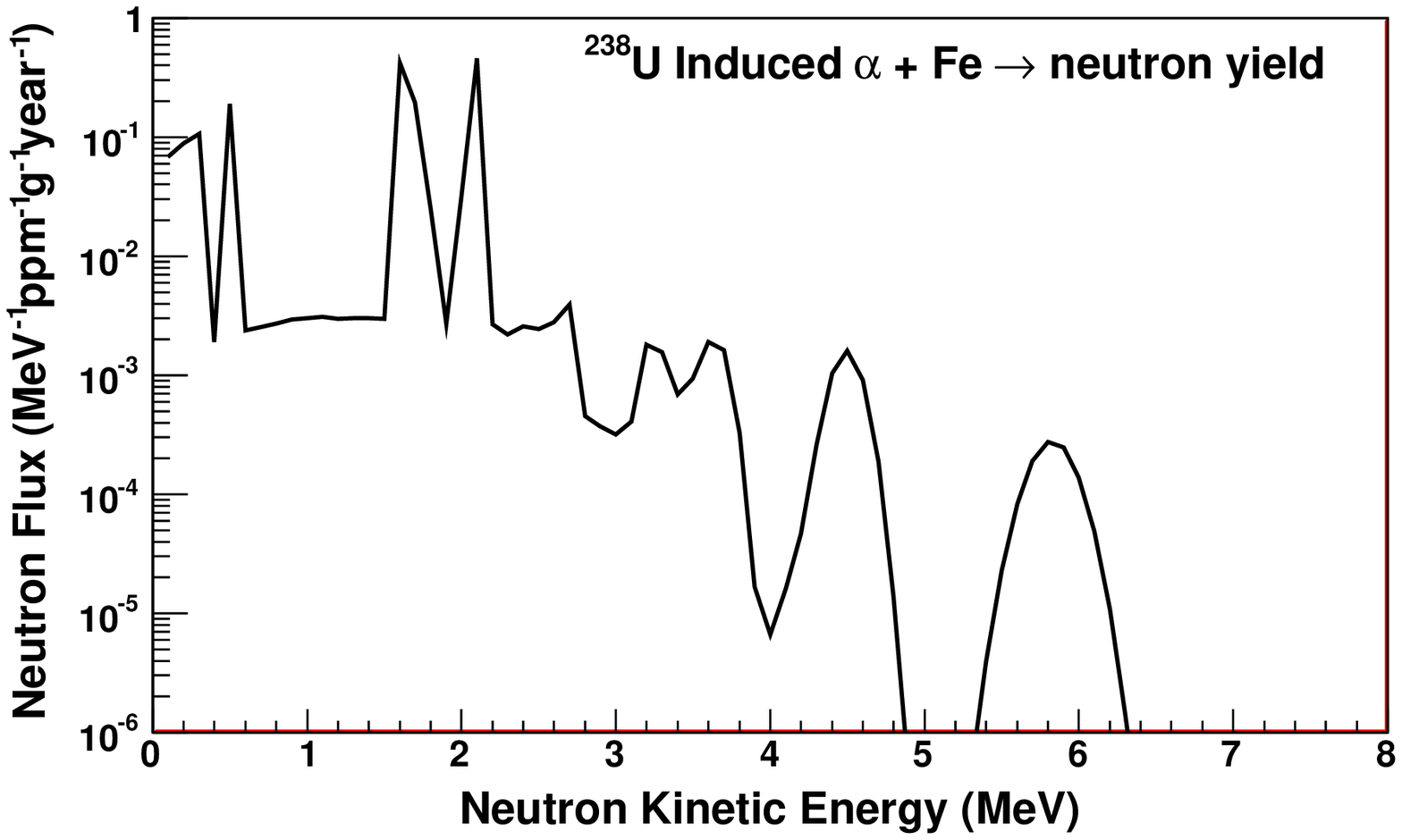}
\includegraphics[width=0.9\textwidth]{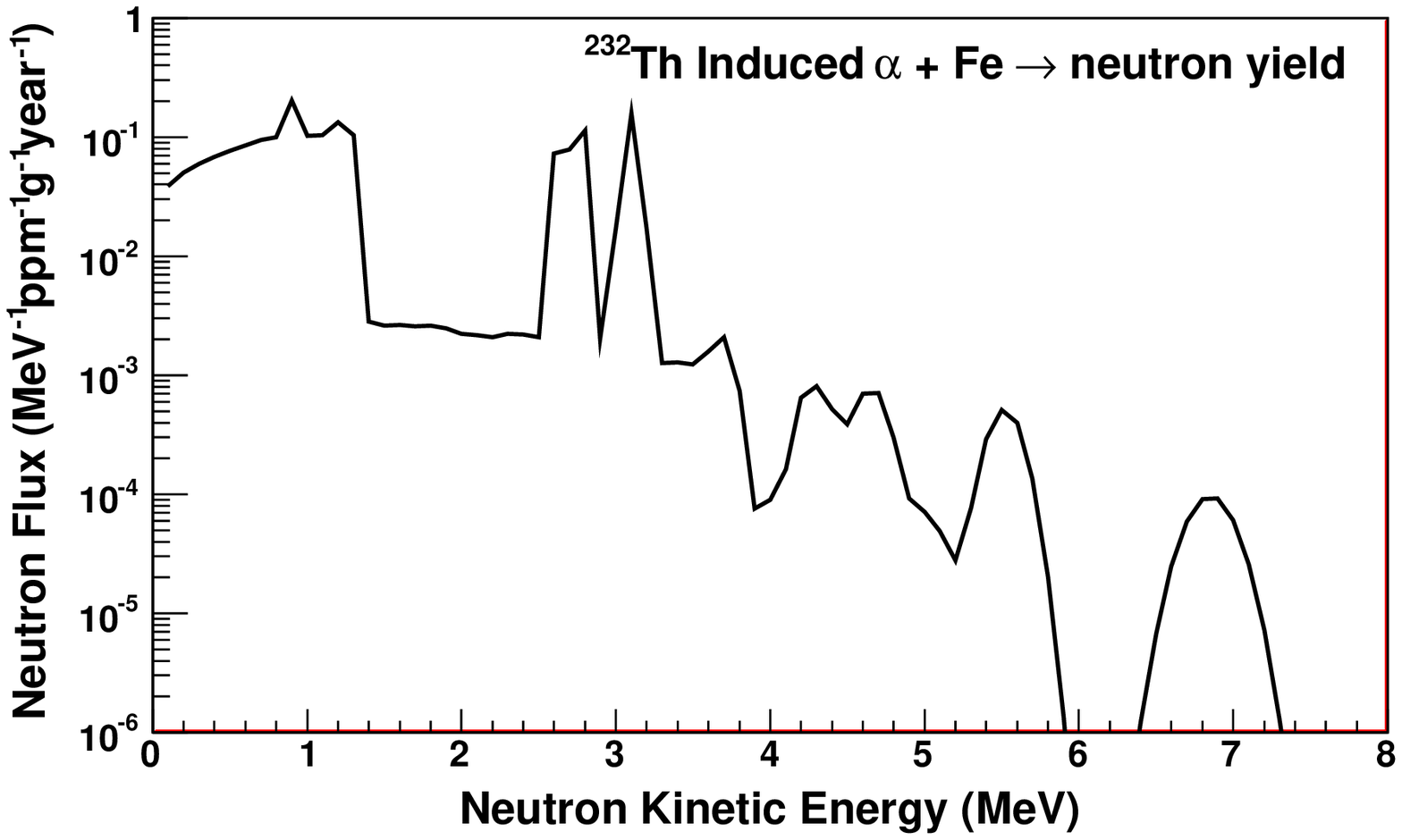}
\caption{The differential neutron flux induced by $(\alpha, n)$ reaction in a thick target of iron. The $\alpha$-particles
  are induced by $^{238}U$ and $^{232}Th$ decays.}\label{fig:Fe}
\end{figure}

The $(\alpha, n)$ induced neutron yield is calculated for a number of elements 
by using Eq.(\ref{diff}) and Eq.(\ref{added}). 
The differential energy spectra of neutrons for specific target elements
are listed in Tables \ref{rateAr} to \ref{rateFe} and plotted in Figures ~\ref{fig:Ar} to ~\ref{fig:Fe}.
 Only the results for neutron energies greater than 0.1 MeV with the 
energy bin size of 0.1 MeV are presented. 
Note that the calculations with the thick target model under equilibrium conditions give 
the maximum contribution of the neutron yield. 
It's worthwhile to note that we also calculated the 
$(\alpha, n)$ neutron yield for lead. However, the result showed that there is no
$(\alpha, n)$ neutron yield in lead due to a very high coulomb barrier which largely restricts the ($\alpha$,n) reactions. 
The neutron yield from spontaneous $^{235}U$ fission is about $~5\%$ of the total neutron yields from uranium and thorium decays. 
Those neutrons from $^{235}U$ with a natural isotropic abundance of $0.72\%$ is not included in these tables.   
The $\alpha$-particles from samarium, a dopant of the glasses for PMTs 
at a level of $\sim$0.1\% to $\sim$1\%~\cite{andr, msio} depending on the type of glasses, have a maximum energy of 2.5 MeV.
As a result, the induced neutrons are less energetic ($<$ 5 MeV) compared to the neutrons caused by uranium and thorium decays.

We compared our results with the calculations made by Heaton {\it et. al.}~\cite{heat1990}, and found the total number of neutron yields
in good agreement. Note that the neutron energy spectra in various elements are quite different. Neutron induced background in the region 
of interest for dark matter experiments is very sensitive to the neutron energy. Thus the neutron energy spectrum
is very important in the Monte Carlo simulation that evaluates the  neutron induced background for various dark matter experiments.

\begin{sidewaystable}
\caption{Neutron production rate via $(\alpha, n)$ reaction due to natural radioactivity. }
\centering
\begin{tabular}{l l|ccccccccccc}\hline
\multicolumn{2}{l}{Range$(MeV)$} &$0\sim .1$ &$.1\sim .2$
                                &$.2\sim .3$ &$.3\sim .4$ &$.4\sim .5$ &$.5\sim .6$
                                &$.6\sim .7$ &$.7\sim .8$ &$.8\sim .9$ &$.9\sim 1$ &Sum \\
\hline
 $E_{n}$ & Source & \multicolumn{11}{c}{$\alpha + Ar \rightarrow $
                         Outgoing Neutron Flux $(ppm^{-1}g^{-1}year^{-1})$}  \\
\hline
&$^{238}U$
&  /   &1.1e-01&1.4e-01&1.7e-01&2.2e-01&2.1e-01&1.9e-01&1.9e-01&1.8e-01&2.1e-01&1.6e+00\\
\raisebox{1.5ex}{$0$}&$^{232}Th$
&  /   &5.4e-02&6.8e-02&7.7e-02&8.6e-02&9.7e-02&9.7e-02&9.0e-02&9.7e-02&9.9e-02&7.6e-01\\
\hline
&$^{238}U$
&2.1e-01&2.0e-01&2.0e-01&1.9e-01&1.9e-01&1.9e-01&2.2e-01&2.0e-01&1.7e-01&1.9e-01&2.0e+00\\
\raisebox{1.5ex}{$1$}&$^{232}Th$
&1.0e-01&1.1e-01&1.0e-01&1.0e-01&1.1e-01&9.2e-02&1.0e-01&1.1e-01&1.1e-01&9.3e-02&1.0e+00\\
\hline
&$^{238}U$
&2.2e-01&2.0e-01&1.6e-01&1.6e-01&1.9e-01&2.3e-01&2.6e-01&1.7e-01&5.1e-02&4.8e-02&1.7e+00\\
\raisebox{1.5ex}{$2$}&$^{232}Th$
&6.9e-02&7.1e-02&6.9e-02&8.2e-02&8.0e-02&5.7e-02&5.9e-02&7.9e-02&7.6e-02&5.5e-02&7.0e-01\\
\hline
&$^{238}U$
&4.9e-02&7.4e-02&7.5e-02&2.6e-02&9.4e-03&4.1e-02&5.7e-02&2.7e-02&1.6e-02&4.5e-02&4.2e-01\\
\raisebox{1.5ex}{$3$}&$^{232}Th$
&5.0e-02&5.6e-02&5.9e-02&5.8e-02&5.4e-02&4.4e-02&4.3e-02&3.8e-02&2.5e-02&1.6e-02&4.4e-01\\
\hline
&$^{238}U$
&6.7e-02&7.0e-02&6.6e-02&4.4e-02&3.1e-02&4.8e-02&5.4e-02&3.1e-02&8.8e-03&1.1e-03&4.2e-01\\
\raisebox{1.5ex}{$4$}&$^{232}Th$
&7.7e-03&8.8e-03&1.1e-02&4.7e-03&2.4e-03&6.4e-03&8.1e-03&4.5e-03&3.1e-03&6.6e-03&6.3e-02\\
\hline
&$^{238}U$
&6.2e-05&1.5e-06&1.5e-08&6.6e-11&1.2e-13&0.0&0.0&0.0&0.0&0.0&6.4e-05\\
\raisebox{1.5ex}{$5$}&$^{232}Th$
&9.8e-03&1.1e-02&1.1e-02&7.7e-03&6.5e-03&9.1e-03&9.7e-03&5.9e-03&2.0e-03&3.4e-04&7.3e-02\\
\hline
&$^{238}U$
&0.0&0.0&0.0&0.0&0.0&0.0&0.0&0.0&0.0&0.0&0.0\\
\raisebox{1.5ex}{$>6$}&$^{232}Th$
&3.0e-05&1.3e-06&2.8e-08&3.0e-10&1.5e-12&3.9e-15&0.0&0.0&0.0&0.0&3.1e-05\\
\hline
&$^{238}U$
&\multicolumn{10}{r}{}&6.1e+00 \\
\raisebox{1.5ex}{Tot}&$^{232}Th$
 &\multicolumn{10}{r}{}&3.1e+00 \\
\hline
\end{tabular}
\label{rateAr}
\end{sidewaystable}

\begin{sidewaystable}
\caption{Neutron production rate via $(\alpha, n)$ reaction due to natural radioactivity. }
\centering
\begin{tabular}{l l|ccccccccccc}\hline
\multicolumn{2}{l}{Range$(MeV)$} &$0\sim .1$ &$.1\sim .2$
                                &$.2\sim .3$ &$.3\sim .4$ &$.4\sim .5$ &$.5\sim .6$
                                &$.6\sim .7$ &$.7\sim .8$ &$.8\sim .9$ &$.9\sim 1$ &Sum \\
\hline
 $E_{n}$ & Source & \multicolumn{11}{c}{$\alpha + Xe \rightarrow $
                         Outgoing Neutron Flux $(ppm^{-1}g^{-1}year^{-1})$}  \\
\hline
&$^{238}U$
 & /       &1.6e-09 &8.4e-10 &0.0 &0.0 &0.0 &0.0 &0.0 &0.0 &0.0 &2.4e-09\\
\raisebox{1.5ex}{$0$}&$^{232}Th$
&  /   &5.8e-08&7.6e-08&9.0e-08&7.7e-08&5.2e-08&3.1e-08&1.9e-08&2.0e-08&2.2e-08&4.4e-07\\
\hline
&$^{238}U$
&0.0&0.0&0.0&0.0&2.9e-10&3.9e-10&3.0e-09&3.5e-09&6.4e-10&4.2e-19&7.9e-09\\
\raisebox{1.5ex}{$1$}&$^{232}Th$
&2.3e-08&2.4e-08&1.7e-08&5.2e-09&0.0&0.0&0.0&0.0&0.0&0.0&6.9e-08\\
\hline
&$^{238}U$
&4.8e-20&3.5e-13&8.3e-10&1.5e-09&6.5e-10&1.7e-13&0.0&0.0&0.0&0.0&3.0e-09\\
\raisebox{1.5ex}{$2$}&$^{232}Th$
&0.0&0.0&2.7e-09&1.2e-08&9.7e-09&3.0e-08&3.0e-08&1.4e-09&6.0e-14&6.1e-13&8.5e-08\\
\hline
&$^{238}U$
 &0.0 &0.0 &0.0 &0.0 &0.0 &0.0 &0.0 &0.0 &0.0 &0.0 &0.0\\
\raisebox{1.5ex}{$>3$}&$^{232}Th$
 &3.3e-10 &3.9e-09 &4.4e-09 &8.0e-10 &3.6e-12 &0.0 &0.0 &0.0 &0.0 &0.0 &9.4e-09\\
\hline
&$^{238}U$
&\multicolumn{10}{r}{}&1.3e-08\\
\raisebox{1.5ex}{Tot}&$^{232}Th$
 &\multicolumn{10}{r}{}&6.1e-07\\
\hline
\end{tabular}
\label{rateXe}
\end{sidewaystable}

\begin{sidewaystable}
\caption{Neutron production rate via $(\alpha, n)$ reaction due to natural radioactivity. }
\centering
\begin{tabular}{l l|ccccccccccc}\hline
\multicolumn{2}{l}{Range$(MeV)$} &$0\sim .1$ &$.1\sim .2$
                                &$.2\sim .3$ &$.3\sim .4$ &$.4\sim .5$ &$.5\sim .6$
                                &$.6\sim .7$ &$.7\sim .8$ &$.8\sim .9$ &$.9\sim 1$ &Sum \\
\hline
 $E_{n}$ & Source & \multicolumn{11}{c}{$\alpha + Ne \rightarrow $
                         Outgoing Neutron Flux $(ppm^{-1}g^{-1}year^{-1})$}  \\
\hline
&$^{238}U$
&  /   &1.1e-01&2.7e-02&3.2e-02&5.2e-02&1.8e-01&2.4e-01&1.3e-01&6.4e-02&8.2e-02&9.1e-01\\
\raisebox{1.5ex}{$0$}&$^{232}Th$
&  /   &3.2e-02&3.8e-02&2.0e-02&1.9e-02&2.0e-02&5.0e-02&4.9e-02&4.3e-02&4.7e-02&3.2e-01\\
\hline
&$^{238}U$
&8.4e-02&1.6e-01&1.6e-01&1.0e-01&1.2e-01&1.3e-01&9.7e-02&8.4e-02&1.2e-01&1.3e-01&1.2e+00\\
\raisebox{1.5ex}{$1$}&$^{232}Th$
&2.2e-02&2.0e-02&3.6e-02&4.3e-02&4.4e-02&4.0e-02&2.1e-02&2.6e-02&3.9e-02&3.4e-02&3.3e-01\\
\hline
&$^{238}U$
&1.4e-01&1.0e-01&5.5e-02&6.2e-02&1.1e-01&1.9e-01&1.8e-01&8.0e-02&7.2e-02&1.1e-01&1.1e+00\\
\raisebox{1.5ex}{$2$}&$^{232}Th$
&3.7e-02&3.7e-02&2.8e-02&3.4e-02&4.4e-02&3.8e-02&3.0e-02&3.0e-02&2.7e-02&2.9e-02&3.3e-01\\
\hline
&$^{238}U$
&1.2e-01&1.2e-01&1.3e-01&1.2e-01&1.2e-01&1.2e-01&9.8e-02&6.9e-02&5.7e-02&7.5e-02&1.0e+00\\
\raisebox{1.5ex}{$3$}&$^{232}Th$
&3.2e-02&3.3e-02&3.5e-02&3.2e-02&3.0e-02&3.7e-02&3.4e-02&2.6e-02&2.1e-02&2.1e-02&3.0e-01\\
\hline
&$^{238}U$
&9.3e-02&8.3e-02&6.6e-02&6.7e-02&6.7e-02&5.2e-02&3.6e-02&2.6e-02&2.3e-02&2.5e-02&5.4e-01\\
\raisebox{1.5ex}{$4$}&$^{232}Th$
&2.7e-02&3.7e-02&3.9e-02&3.1e-02&2.1e-02&1.6e-02&1.6e-02&1.7e-02&1.7e-02&1.7e-02&2.4e-01\\
\hline
&$^{238}U$
&2.6e-02&2.1e-02&1.5e-02&1.3e-02&1.2e-02&1.1e-02&9.4e-03&1.2e-02&1.9e-02&2.6e-02&1.7e-01\\
\raisebox{1.5ex}{$5$}&$^{232}Th$
&1.6e-02&1.6e-02&1.7e-02&1.7e-02&1.1e-02&5.3e-03&2.4e-03&2.3e-03&3.5e-03&4.4e-03&9.5e-02\\
\hline
&$^{238}U$
&2.8e-02&2.3e-02&1.4e-02&6.8e-03&2.8e-03&1.2e-03&7.3e-04&5.6e-04&4.9e-04&4.6e-04&7.8e-02\\
\raisebox{1.5ex}{$6$}&$^{232}Th$
&4.1e-03&3.1e-03&2.4e-03&2.2e-03&2.0e-03&1.8e-03&2.1e-03&3.2e-03&4.6e-03&5.2e-03&3.1e-02\\
\hline
&$^{238}U$
&4.6e-04&4.9e-04&5.6e-04&6.6e-04&7.6e-04&8.3e-04&8.3e-04&7.4e-04&5.8e-04&4.1e-04&6.3e-03\\
\raisebox{1.5ex}{$7$}&$^{232}Th$
&4.5e-03&3.0e-03&1.6e-03&6.6e-04&2.7e-04&1.4e-04&1.1e-04&9.7e-05&9.2e-05&9.1e-05&1.1e-02\\
\hline
&$^{238}U$
&2.5e-04&1.4e-04&7.8e-05&4.6e-05&3.8e-05&4.3e-05&5.6e-05&7.2e-05&8.6e-05&9.5e-05&9.1e-04\\
\raisebox{1.5ex}{$8$}&$^{232}Th$
&9.3e-05&1.0e-04&1.1e-04&1.3e-04&1.4e-04&1.4e-04&1.3e-04&9.9e-05&6.7e-05&4.0e-05&1.0e-03\\
\hline
&$^{238}U$
&9.6e-05&8.8e-05&7.4e-05&5.7e-05&4.0e-05&2.6e-05&1.5e-05&8.3e-06&4.1e-06&1.8e-06&4.1e-04\\
\raisebox{1.5ex}{$>9$}&$^{232}Th$
&2.1e-05&9.9e-06&5.1e-06&4.2e-06&5.4e-06&7.8e-06&1.1e-05&1.4e-05&1.5e-05&1.6e-05&1.1e-04\\
\hline
&$^{238}U$
&\multicolumn{10}{r}{}&5.0e+00\\
\raisebox{1.5ex}{Tot}&$^{232}Th$
 &\multicolumn{10}{r}{}&1.7e+00\\
\hline
\end{tabular}
\label{rateNe}
\end{sidewaystable}

\begin{sidewaystable}
\caption{Neutron production rate via $(\alpha, n)$ reaction due to natural radioactivity. }
\centering
\begin{tabular}{l l|ccccccccccc}\hline
\multicolumn{2}{l}{Range$(MeV)$} &$0\sim .1$ &$.1\sim .2$
                                &$.2\sim .3$ &$.3\sim .4$ &$.4\sim .5$ &$.5\sim .6$
                                &$.6\sim .7$ &$.7\sim .8$ &$.8\sim .9$ &$.9\sim 1$ &Sum \\
\hline
 $E_{n}$ & Source & \multicolumn{10}{c}{$\alpha + Cu \rightarrow $
                         Outgoing Neutron Flux $(ppm^{-1}g^{-1}year^{-1})$}  \\
\hline
&$^{238}U$
&  /   &1.1e-03&1.5e-03&1.7e-03&2.0e-03&2.1e-03&2.2e-03&1.1e-03&3.4e-08&1.5e-08&1.2e-02\\
\raisebox{1.5ex}{$0$}&$^{232}Th$
&  /   &3.8e-03&5.0e-03&3.7e-03&1.8e-03&2.0e-03&2.1e-03&2.1e-03&2.1e-03&2.1e-03&2.5e-02\\
\hline
&$^{238}U$
&5.7e-09&1.8e-09&3.5e-10&2.2e-11&1.2e-08&1.2e-08&0.0&0.0&0.0&0.0&3.2e-08\\
\raisebox{1.5ex}{$1$}&$^{232}Th$
&2.0e-03&1.9e-03&1.8e-03&1.7e-03&1.6e-03&1.5e-03&1.4e-03&7.0e-04&1.5e-06&4.6e-09&1.3e-02\\
\hline
&$^{238}U$
 &0.0 &0.0 &0.0 &0.0 &0.0 &0.0 &0.0 &0.0 &0.0 &0.0 &0.0\\
\raisebox{1.5ex}{$>2$}&$^{232}Th$
&2.0e-09&8.6e-10&2.3e-10&7.8e-11&1.1e-04&1.1e-04&7.0e-09&0.0&0.0&0.0&2.3e-04\\
\hline
&$^{238}U$
&\multicolumn{10}{r}{}&1.2e-02\\
\raisebox{1.5ex}{Tot}&$^{232}Th$
 &\multicolumn{10}{r}{}&3.8e-02\\
\hline
\end{tabular}
\label{rateCu}
\end{sidewaystable}

\begin{sidewaystable}
\caption{Neutron production rate via $(\alpha, n)$ reaction due to natural radioactivity. }
\centering
\begin{tabular}{l l|ccccccccccc}\hline
\multicolumn{2}{l}{Range$(MeV)$} &$0\sim .1$ &$.1\sim .2$
                                &$.2\sim .3$ &$.3\sim .4$ &$.4\sim .5$ &$.5\sim .6$
                                &$.6\sim .7$ &$.7\sim .8$ &$.8\sim .9$ &$.9\sim 1$ &Sum \\
\hline
 $E_{n}$ & Source & \multicolumn{11}{c}{$\alpha + B \rightarrow $
                         Outgoing Neutron Flux $(ppm^{-1}g^{-1}year^{-1})$}  \\
\hline
&$^{238}U$
&  /   &1.0e-01&1.0e-01&1.2e-06&1.6e-01&1.7e-01&8.8e-03&8.9e-02&6.6e-01&6.6e-01&2.0e+00\\
\raisebox{1.5ex}{$0$}&$^{232}Th$
&  /   &3.1e-07&1.7e-06&9.7e-03&9.7e-03&1.1e-01&1.6e-01&8.2e-02&3.0e-02&1.0e-06&4.1e-01\\
\hline
&$^{238}U$
&9.6e-02&2.4e-01&9.8e-01&9.9e-01&3.7e-01&2.0e-01&4.0e-01&1.7e+00&1.6e+00&2.8e-01&6.8e+00\\
\raisebox{1.5ex}{$1$}&$^{232}Th$
&7.6e-02&8.3e-02&2.9e-02&8.9e-02&9.7e-02&1.0e-01&1.3e-01&1.4e-01&1.7e-01&2.1e-01&1.1e+00\\
\hline
&$^{238}U$
&1.7e-01&3.8e-01&3.6e-01&1.3e-01&5.5e-02&7.0e-02&8.2e-02&1.3e-01&3.4e-01&7.4e-01&2.5e+00\\
\raisebox{1.5ex}{$2$}&$^{232}Th$
&1.7e-01&8.4e-02&1.3e-01&2.3e-01&2.1e-01&2.6e-01&3.0e-01&2.5e-01&2.5e-01&2.4e-01&2.1e+00\\
\hline
&$^{238}U$
&1.1e+00&1.2e+00&1.5e+00&2.1e+00&2.7e+00&2.9e+00&2.8e+00&2.5e+00&2.3e+00&2.1e+00&2.1e+01\\
\raisebox{1.5ex}{$3$}&$^{232}Th$
&2.4e-01&2.2e-01&1.6e-01&1.0e-01&6.1e-02&5.3e-02&9.2e-02&1.8e-01&2.8e-01&3.8e-01&1.8e+00\\
\hline
&$^{238}U$
&2.1e+00&2.0e+00&1.8e+00&1.6e+00&1.4e+00&1.2e+00&9.3e-01&6.3e-01&3.8e-01&2.1e-01&1.2e+01\\
\raisebox{1.5ex}{$4$}&$^{232}Th$
&4.9e-01&6.0e-01&6.3e-01&5.9e-01&5.3e-01&5.0e-01&5.0e-01&5.0e-01&5.0e-01&4.8e-01&5.3e+00\\
\hline
&$^{238}U$
&1.3e-01&9.9e-02&1.2e-01&2.0e-01&3.6e-01&5.9e-01&8.1e-01&9.0e-01&8.1e-01&5.9e-01&4.6e+00\\
\raisebox{1.5ex}{$5$}&$^{232}Th$
&4.4e-01&3.7e-01&2.7e-01&1.7e-01&9.6e-02&5.1e-02&3.0e-02&2.2e-02&1.8e-02&1.7e-02&1.5e+00\\
\hline
&$^{238}U$
&3.4e-01&1.7e-01&7.4e-02&4.1e-02&3.3e-02&3.2e-02&2.9e-02&2.3e-02&1.6e-02&1.0e-02&7.7e-01\\
\raisebox{1.5ex}{$6$}&$^{232}Th$
&2.1e-02&3.5e-02&6.6e-02&1.1e-01&1.6e-01&1.8e-01&1.6e-01&1.2e-01&6.8e-02&3.3e-02&9.4e-01\\
\hline
&$^{238}U$
&5.2e-03&2.4e-03&9.5e-04&3.3e-04&9.7e-05&2.5e-05&5.6e-06&1.1e-06&1.8e-07&2.6e-08&9.1e-03\\
\raisebox{1.5ex}{$7$}&$^{232}Th$
&1.5e-02&7.8e-03&6.2e-03&5.8e-03&5.1e-03&4.0e-03&2.6e-03&1.5e-03&7.5e-04&3.2e-04&4.9e-02\\
\hline
&$^{238}U$
&3.3e-09&3.5e-10&3.3e-11&2.7e-12&1.9e-13&1.1e-14&0.0&0.0&0.0&0.0&3.7e-09\\
\raisebox{1.5ex}{$>8$}&$^{232}Th$
&1.2e-04&3.7e-05&9.9e-06&2.3e-06&4.5e-07&7.7e-08&1.1e-08&1.4e-09&1.5e-10&1.4e-11&1.7e-04\\
\hline
&$^{238}U$
&\multicolumn{10}{r}{}&5.0e+01\\
\raisebox{1.5ex}{Tot}&$^{232}Th$
 &\multicolumn{10}{r}{}&1.3e+01\\
\hline
\end{tabular}
\label{rateB}
\end{sidewaystable}

\begin{sidewaystable}
\caption{Neutron production rate via $(\alpha, n)$ reaction due to natural radioactivity. }
\centering
\begin{tabular}{l l|ccccccccccc}\hline
\multicolumn{2}{l}{Range$(MeV)$} &$0\sim .1$ &$.1\sim .2$
                                &$.2\sim .3$ &$.3\sim .4$ &$.4\sim .5$ &$.5\sim .6$
                                &$.6\sim .7$ &$.7\sim .8$ &$.8\sim .9$ &$.9\sim 1$ &Sum\\
\hline
 $E_{n}$ & Source & \multicolumn{11}{c}{$\alpha + C \rightarrow $
                         Outgoing Neutron Flux $(ppm^{-1}g^{-1}year^{-1})$}  \\
\hline
&$^{238}U$
&  /   &9.3e-10&5.0e-06&5.0e-06&8.2e-10&7.6e-10&6.2e-03&6.3e-03&4.8e-05&3.2e-10&1.3e-02\\
\raisebox{1.5ex}{$0$}&$^{232}Th$
&  /   &7.2e-10&2.2e-05&2.2e-05&9.5e-04&9.5e-04&3.2e-03&4.2e-03&1.8e-03&1.4e-03&1.3e-02\\
\hline
&$^{238}U$
&5.0e-11&1.7e-14&2.3e-07&2.1e-06&1.9e-06&8.7e-08&1.3e-08&4.7e-06&9.1e-03&3.4e-02&4.3e-02\\
\raisebox{1.5ex}{$1$}&$^{232}Th$
&5.9e-04&2.7e-04&7.2e-03&7.3e-03&3.3e-04&5.7e-10&7.2e-09&6.4e-08&1.1e-07&1.0e-07&1.6e-02\\
\hline
&$^{238}U$
&2.7e-02&1.7e-03&3.3e-06&2.6e-07&2.5e-07&2.3e-07&2.2e-07&2.1e-07&2.0e-07&1.8e-07&2.8e-02\\
\raisebox{1.5ex}{$2$}&$^{232}Th$
&9.5e-08&9.2e-08&8.8e-08&8.4e-08&8.2e-08&5.1e-06&9.5e-04&6.8e-03&7.5e-03&1.7e-03&1.7e-02\\
\hline
&$^{238}U$
&1.7e-07&1.7e-07&1.7e-07&2.0e-07&2.9e-07&5.4e-07&1.2e-06&2.7e-06&6.0e-06&1.3e-05&2.4e-05\\
\raisebox{1.5ex}{$3$}&$^{232}Th$
&7.3e-05&4.1e-07&1.0e-07&1.8e-07&3.6e-07&7.7e-07&1.7e-06&3.5e-06&7.0e-06&1.3e-05&1.0e-04\\
\hline
&$^{238}U$
&2.7e-05&5.4e-05&1.0e-04&1.8e-04&3.2e-04&5.4e-04&8.6e-04&1.3e-03&2.0e-03&2.9e-03&8.3e-03\\
\raisebox{1.5ex}{$4$}&$^{232}Th$
&2.5e-05&4.4e-05&7.3e-05&1.2e-04&1.8e-04&2.6e-04&3.6e-04&4.7e-04&5.9e-04&7.1e-04&2.8e-03\\
\hline
&$^{238}U$
&4.0e-03&5.4e-03&7.1e-03&8.9e-03&1.1e-02&1.3e-02&1.4e-02&1.6e-02&1.7e-02&1.7e-02&1.1e-01\\
\raisebox{1.5ex}{$5$}&$^{232}Th$
&8.1e-04&8.9e-04&9.4e-04&9.6e-04&9.6e-04&9.6e-04&1.0e-03&1.1e-03&1.3e-03&1.5e-03&1.0e-02\\
\hline
&$^{238}U$
&1.8e-02&1.7e-02&1.7e-02&1.6e-02&1.4e-02&1.3e-02&1.2e-02&1.0e-02&8.3e-03&6.7e-03&1.3e-01\\
\raisebox{1.5ex}{$6$}&$^{232}Th$
&1.9e-03&2.3e-03&2.7e-03&3.0e-03&3.3e-03&3.5e-03&3.7e-03&3.7e-03&3.6e-03&3.5e-03&3.1e-02\\
\hline
&$^{238}U$
&5.2e-03&3.8e-03&2.7e-03&1.9e-03&1.5e-03&1.4e-03&1.6e-03&2.0e-03&2.4e-03&2.9e-03&2.5e-02\\
\raisebox{1.5ex}{$7$}&$^{232}Th$
&3.3e-03&3.1e-03&2.8e-03&2.5e-03&2.1e-03&1.7e-03&1.3e-03&9.4e-04&6.3e-04&4.0e-04&1.9e-02\\
\hline
&$^{238}U$
&3.1e-03&3.1e-03&2.8e-03&2.3e-03&1.7e-03&1.2e-03&7.3e-04&4.2e-04&2.1e-04&1.0e-04&1.6e-02\\
\raisebox{1.5ex}{$8$}&$^{232}Th$
&2.4e-04&1.5e-04&1.1e-04&1.2e-04&1.8e-04&2.6e-04&3.7e-04&4.6e-04&5.3e-04&5.5e-04&3.0e-03\\
\hline
&$^{238}U$
&4.3e-05&1.7e-05&6.0e-06&1.9e-06&5.7e-07&1.5e-07&3.7e-08&8.2e-09&1.7e-09&3.0e-10&6.9e-05\\
\raisebox{1.5ex}{$>9$}&$^{232}Th$
&5.1e-04&4.3e-04&3.2e-04&2.2e-04&1.4e-04&7.6e-05&3.8e-05&1.7e-05&7.0e-06&2.6e-06&1.8e-03\\
\hline
&$^{238}U$
&\multicolumn{10}{r}{}&3.8e-01\\
\raisebox{1.5ex}{Tot}&$^{232}Th$
 &\multicolumn{10}{r}{}&1.1e-01\\
\hline
\end{tabular}
\label{rateC}
\end{sidewaystable}

\begin{sidewaystable}
\caption{Neutron production rate via $(\alpha, n)$ reaction due to natural radioactivity. }
\centering
\begin{tabular}{l l|ccccccccccc}\hline
\multicolumn{2}{l}{Range$(MeV)$} &$0\sim .1$ &$.1\sim .2$
                                &$.2\sim .3$ &$.3\sim .4$ &$.4\sim .5$ &$.5\sim .6$
                                &$.6\sim .7$ &$.7\sim .8$ &$.8\sim .9$ &$.9\sim 1$ &Sum\\
\hline
$E_{n}$ & Source & \multicolumn{11}{c}{$\alpha + O \rightarrow $
                         Outgoing Neutron Flux $(ppm^{-1}g^{-1}year^{-1})$}  \\
\hline
&$^{238}U$
&  /   &2.2e-04&2.6e-03&2.7e-03&4.8e-04&4.1e-04&6.4e-04&1.2e-03&1.9e-03&4.3e-03&1.4e-02\\
\raisebox{1.5ex}{$0$}&$^{232}Th$
&  /   &1.3e-04&1.3e-04&2.6e-04&3.7e-04&4.8e-04&3.5e-04&4.2e-04&7.5e-04&6.0e-04&3.5e-03\\
\hline
&$^{238}U$
&3.1e-03&6.4e-06&1.3e-03&4.3e-03&4.4e-03&1.3e-03&1.8e-03&2.4e-03&3.2e-03&3.8e-03&2.6e-02\\
\raisebox{1.5ex}{$1$}&$^{232}Th$
&1.3e-03&2.0e-03&9.3e-04&3.9e-04&7.3e-04&1.0e-03&9.4e-04&4.5e-04&6.5e-04&1.6e-03&1.0e-02\\
\hline
&$^{238}U$
&2.0e-03&1.1e-03&8.6e-04&2.1e-03&3.2e-03&3.5e-03&5.7e-03&8.3e-03&7.3e-03&5.2e-03&3.9e-02\\
\raisebox{1.5ex}{$2$}&$^{232}Th$
&1.6e-03&1.1e-03&8.8e-04&5.2e-04&5.2e-04&9.6e-04&1.2e-03&1.0e-03&6.1e-04&3.2e-04&8.7e-03\\
\hline
&$^{238}U$
&4.5e-03&5.0e-03&5.6e-03&5.3e-03&4.4e-03&3.7e-03&3.9e-03&5.0e-03&5.5e-03&4.0e-03&4.7e-02\\
\raisebox{1.5ex}{$3$}&$^{232}Th$
&6.5e-04&7.9e-04&6.8e-04&7.8e-04&1.1e-03&1.6e-03&1.8e-03&1.5e-03&1.2e-03&1.2e-03&1.1e-02\\
\hline
&$^{238}U$
&2.3e-03&1.6e-03&1.3e-03&9.3e-04&5.3e-04&2.7e-04&1.5e-04&1.3e-04&1.8e-04&4.0e-04&7.8e-03\\
\raisebox{1.5ex}{$4$}&$^{232}Th$
&1.1e-03&1.0e-03&9.4e-04&9.9e-04&1.1e-03&1.1e-03&1.0e-03&9.2e-04&6.9e-04&4.2e-04&9.3e-03\\
\hline
&$^{238}U$
&8.8e-04&1.5e-03&1.8e-03&1.7e-03&1.5e-03&1.2e-03&1.0e-03&6.7e-04&3.4e-04&1.3e-04&1.1e-02\\
\raisebox{1.5ex}{$5$}&$^{232}Th$
&2.2e-04&1.1e-04&5.0e-05&3.1e-05&2.6e-05&2.5e-05&2.8e-05&4.2e-05&8.4e-05&1.7e-04&7.8e-04\\
\hline
&$^{238}U$
&3.6e-05&8.0e-06&2.6e-06&3.5e-06&7.8e-06&1.6e-05&2.6e-05&3.8e-05&4.5e-05&4.5e-05&2.3e-04\\
\raisebox{1.5ex}{$6$}&$^{232}Th$
&2.7e-04&3.2e-04&3.0e-04&2.5e-04&2.1e-04&1.6e-04&1.1e-04&5.4e-05&2.0e-05&5.6e-06&1.7e-03\\
\hline
&$^{238}U$
&3.8e-05&2.6e-05&1.6e-05&7.8e-06&3.2e-06&1.1e-06&3.3e-07&8.0e-08&1.6e-08&2.8e-09&9.2e-05\\
\raisebox{1.5ex}{$7$}&$^{232}Th$
&1.2e-06&3.6e-07&6.0e-07&1.4e-06&2.9e-06&4.8e-06&6.8e-06&7.9e-06&7.8e-06&6.3e-06&4.0e-05\\
\hline
&$^{238}U$
&4.0e-10&4.8e-11&4.8e-12&4.0e-13&2.8e-14&1.6e-15&0.0&0.0&0.0&0.0&4.5e-10\\
\raisebox{1.5ex}{$>8$}&$^{232}Th$
&4.3e-06&2.4e-06&1.2e-06&4.5e-07&1.5e-07&4.1e-08&9.2e-09&1.7e-09&2.7e-10&3.5e-11&8.6e-06\\
\hline
&$^{238}U$
&\multicolumn{10}{r}{}&1.5e-01\\
\raisebox{1.5ex}{Tot}&$^{232}Th$
 &\multicolumn{10}{r}{}&4.5e-02\\
\hline
\end{tabular}
\label{rateO}
\end{sidewaystable}

\begin{sidewaystable}
\caption{Neutron production rate via $(\alpha, n)$ reaction due to natural radioactivity. }
\centering
\begin{tabular}{l l|ccccccccccc}\hline
\multicolumn{2}{l}{Range$(MeV)$} &$0\sim .1$ &$.1\sim .2$
                                &$.2\sim .3$ &$.3\sim .4$ &$.4\sim .5$ &$.5\sim .6$
                                &$.6\sim .7$ &$.7\sim .8$ &$.8\sim .9$ &$.9\sim 1$ &Sum\\
\hline
 $E_{n}$ & Source & \multicolumn{11}{c}{$\alpha + Al \rightarrow $
                         Outgoing Neutron Flux $(ppm^{-1}g^{-1}year^{-1})$}  \\
\hline
&$^{238}U$
&  /   &3.1e-02&7.7e-02&5.6e-02&7.6e-03&3.3e-01&4.3e-01&2.0e-01&8.7e-02&1.9e-02&1.2e+00\\
\raisebox{1.5ex}{$0$}&$^{232}Th$
&  /   &8.0e-03&1.2e-02&1.6e-02&1.8e-02&2.3e-02&4.4e-02&1.1e-01&1.6e-01&9.2e-02&4.9e-01\\
\hline
&$^{238}U$
&1.1e-01&9.5e-02&6.4e-02&1.5e-01&5.0e-01&4.3e-01&1.4e-02&1.9e-02&1.3e-01&2.0e-01&1.7e+00\\
\raisebox{1.5ex}{$1$}&$^{232}Th$
&2.3e-02&3.9e-02&1.6e-01&1.7e-01&4.0e-02&3.0e-02&3.1e-02&6.5e-02&7.4e-02&3.1e-02&6.6e-01\\
\hline
&$^{238}U$
&3.8e-01&3.3e-01&3.8e-02&3.7e-03&5.7e-02&2.8e-01&2.6e-01&3.3e-02&5.6e-04&2.8e-06&1.4e+00\\
\raisebox{1.5ex}{$2$}&$^{232}Th$
&5.6e-02&5.8e-02&4.3e-02&2.8e-02&6.4e-02&8.8e-02&4.5e-02&3.1e-02&3.3e-02&5.9e-02&5.1e-01\\
\hline
&$^{238}U$
&2.1e-04&1.1e-02&8.2e-02&1.4e-01&8.1e-02&1.1e-02&3.0e-04&1.9e-03&2.0e-02&7.3e-02&4.2e-01\\
\raisebox{1.5ex}{$3$}&$^{232}Th$
&6.7e-02&4.0e-02&3.6e-02&2.6e-02&2.7e-02&4.0e-02&2.2e-02&1.7e-03&1.2e-05&2.7e-05&2.6e-01\\
\hline
&$^{238}U$
&1.1e-01&7.4e-02&2.0e-02&2.0e-03&6.6e-05&7.3e-07&2.6e-09&3.0e-12&1.1e-15&0.0&2.1e-01\\
\raisebox{1.5ex}{$4$}&$^{232}Th$
&9.3e-04&8.4e-03&2.3e-02&2.4e-02&9.7e-03&1.2e-03&2.2e-04&1.8e-03&8.4e-03&1.8e-02&9.6e-02\\
\hline
&$^{238}U$
 &0.0 &0.0 &0.0 &0.0 &0.0 &0.0 &0.0 &0.0 &0.0 &0.0 &0.0\\
\raisebox{1.5ex}{$>5$}&$^{232}Th$
&1.8e-02&9.5e-03&2.3e-03&2.4e-04&1.1e-05&2.1e-07&1.6e-09&5.1e-12&6.7e-15&0.0&3.0e-02\\
\hline
&$^{238}U$
&\multicolumn{10}{r}{}&5.0e+00\\
\raisebox{1.5ex}{Tot}&$^{232}Th$
 &\multicolumn{10}{r}{}&2.0e+00\\
\hline
\end{tabular}
\label{rateAl}
\end{sidewaystable}

\begin{sidewaystable}
\caption{Neutron production rate via $(\alpha, n)$ reaction due to natural radioactivity. }
\centering
\begin{tabular}{l l|ccccccccccc}\hline
\multicolumn{2}{l}{Range$(MeV)$} &$0\sim .1$ &$.1\sim .2$
                                &$.2\sim .3$ &$.3\sim .4$ &$.4\sim .5$ &$.5\sim .6$
                                &$.6\sim .7$ &$.7\sim .8$ &$.8\sim .9$ &$.9\sim 1$  &Sum \\
\hline
 $E_{n}$ & Source & \multicolumn{11}{c}{$\alpha + Si \rightarrow $
                         Outgoing Neutron Flux $(ppm^{-1}g^{-1}year^{-1})$}  \\
\hline
&$^{238}U$
&  /   &2.7e-05&5.1e-03&8.9e-03&1.4e-02&1.2e-02&1.0e-02&8.8e-03&2.8e-02&2.9e-02&1.2e-01\\
\raisebox{1.5ex}{$0$}&$^{232}Th$
&  /   &8.6e-04&5.5e-04&1.2e-06&1.2e-06&1.2e-06&9.1e-04&9.1e-04&3.7e-04&8.2e-03&1.2e-02\\
\hline
&$^{238}U$
&9.4e-03&2.5e-02&2.1e-02&9.4e-03&3.3e-02&2.9e-02&1.8e-03&2.4e-02&2.4e-02&1.4e-03&1.8e-01\\
\raisebox{1.5ex}{$1$}&$^{232}Th$
&7.9e-03&4.8e-03&1.0e-02&5.5e-03&5.9e-03&9.7e-03&9.3e-03&1.2e-02&8.8e-03&6.6e-03&8.1e-02\\
\hline
&$^{238}U$
&1.4e-03&2.2e-03&1.6e-03&5.5e-03&1.1e-02&1.2e-02&1.0e-02&7.1e-03&9.0e-03&3.2e-02&9.3e-02\\
\raisebox{1.5ex}{$2$}&$^{232}Th$
&9.5e-03&1.6e-02&1.2e-02&2.8e-03&8.6e-03&7.4e-03&7.1e-04&4.2e-06&2.8e-05&2.2e-04&5.8e-02\\
\hline
&$^{238}U$
&3.5e-02&2.1e-02&3.2e-02&2.6e-02&7.8e-03&3.4e-03&6.0e-03&7.8e-03&5.7e-03&2.3e-03&1.5e-01\\
\raisebox{1.5ex}{$3$}&$^{232}Th$
&8.6e-04&1.9e-03&2.7e-03&3.8e-03&4.0e-03&2.4e-03&1.4e-03&1.9e-03&5.0e-03&8.1e-03&3.2e-02\\
\hline
&$^{238}U$
&5.1e-04&5.9e-05&3.6e-06&4.1e-07&6.2e-07&8.3e-07&2.4e-06&2.5e-05&2.0e-04&9.6e-04&1.8e-03\\
\raisebox{1.5ex}{$4$}&$^{232}Th$
&6.3e-03&4.4e-03&5.3e-03&4.6e-03&3.0e-03&1.8e-03&7.2e-04&1.7e-04&2.3e-05&1.7e-06&2.6e-02\\
\hline
&$^{238}U$
&2.8e-03&5.1e-03&5.8e-03&4.2e-03&1.8e-03&5.0e-04&8.3e-05&8.2e-06&4.8e-07&1.6e-08&2.0e-02\\
\raisebox{1.5ex}{$5$}&$^{232}Th$
&7.2e-08&2.8e-09&9.7e-10&7.9e-10&4.8e-09&1.0e-07&1.4e-06&1.2e-05&6.9e-05&2.4e-04&3.3e-04\\
\hline
&$^{238}U$
 &3.3e-10 &3.9e-12 &2.7e-14 &0.0 &0.0 &0.0 &0.0 &0.0 &0.0 &0.0 &3.4e-10\\
\raisebox{1.5ex}{$6$}&$^{232}Th$
&5.6e-04&8.2e-04&7.9e-04&5.0e-04&2.0e-04&5.3e-05&8.8e-06&9.1e-07&6.0e-08&2.4e-09&2.9e-03\\
\hline
&$^{238}U$
 &0.0 &0.0 &0.0 &0.0 &0.0 &0.0 &0.0 &0.0 &0.0 &0.0 &0.0\\
\raisebox{1.5ex}{$>7$}&$^{232}Th$
 &6.1e-11 &9.5e-13 &9.1e-15 &0.0 &0.0 &0.0 &0.0 &0.0 &0.0 &0.0 &6.2e-11\\
\hline
&$^{238}U$
&\multicolumn{10}{r}{}&5.5e-01\\
\raisebox{1.5ex}{Tot}&$^{232}Th$
 &\multicolumn{10}{r}{}&2.1e-01\\
\hline
\end{tabular}
\label{rateSi}
\end{sidewaystable}

\begin{sidewaystable}
\caption{Neutron production rate via $(\alpha, n)$ reaction due to natural radioactivity. }
\centering
\begin{tabular}{l l|ccccccccccc}\hline
\multicolumn{2}{l}{Range$(MeV)$} &$0\sim .1$ &$.1\sim .2$
                                &$.2\sim .3$ &$.3\sim .4$ &$.4\sim .5$ &$.5\sim .6$
                                &$.6\sim .7$ &$.7\sim .8$ &$.8\sim .9$ &$.9\sim 1$  &Sum \\
\hline
 $E_{n}$ & Source & \multicolumn{11}{c}{$\alpha + Fe \rightarrow $
                         Outgoing Neutron Flux $(ppm^{-1}g^{-1}year^{-1})$}  \\
\hline
&$^{238}U$
&  /   &7.8e-03&9.7e-03&5.4e-03&9.6e-03&9.6e-03&2.5e-04&2.6e-04&2.8e-04&3.0e-04&4.3e-02\\
\raisebox{1.5ex}{$0$}&$^{232}Th$
&  /   &4.4e-03&5.5e-03&6.4e-03&7.3e-03&8.1e-03&9.0e-03&9.7e-03&1.5e-02&1.5e-02&8.1e-02\\
\hline
&$^{238}U$
&3.1e-04&3.0e-04&3.0e-04&3.0e-04&3.0e-04&2.1e-02&3.1e-02&1.1e-02&1.4e-03&1.7e-03&6.7e-02\\
\raisebox{1.5ex}{$1$}&$^{232}Th$
&1.0e-02&1.2e-02&1.2e-02&5.3e-03&2.7e-04&2.6e-04&2.6e-04&2.6e-04&2.5e-04&2.4e-04&4.1e-02\\
\hline
&$^{238}U$
&2.4e-02&2.3e-02&2.4e-04&2.4e-04&2.5e-04&2.6e-04&3.3e-04&2.2e-04&4.1e-05&3.4e-05&4.9e-02\\
\raisebox{1.5ex}{$2$}&$^{232}Th$
&2.2e-04&2.1e-04&2.1e-04&2.2e-04&2.1e-04&3.7e-03&7.6e-03&9.6e-03&5.8e-03&9.4e-04&2.9e-02\\
\hline
&$^{238}U$
&3.6e-05&1.1e-04&1.7e-04&1.1e-04&8.1e-05&1.4e-04&1.8e-04&9.7e-05&1.7e-05&1.2e-06&9.4e-04\\
\raisebox{1.5ex}{$3$}&$^{232}Th$
&8.8e-03&8.8e-03&9.4e-04&1.3e-04&1.3e-04&1.4e-04&1.8e-04&1.4e-04&4.1e-05&8.3e-06&1.9e-02\\
\hline
&$^{238}U$
&1.1e-06&3.1e-06&1.5e-05&6.5e-05&1.3e-04&1.3e-04&5.5e-05&1.0e-05&7.1e-07&2.0e-08&4.1e-04\\
\raisebox{1.5ex}{$4$}&$^{232}Th$
&1.3e-05&4.1e-05&7.3e-05&6.6e-05&4.5e-05&5.4e-05&7.0e-05&5.0e-05&2.0e-05&8.2e-06&4.4e-04\\
\hline
&$^{238}U$
&2.3e-09&3.5e-09&2.4e-08&2.2e-07&1.3e-06&5.3e-06&1.4e-05&2.3e-05&2.6e-05&1.9e-05&8.9e-05\\
\raisebox{1.5ex}{$5$}&$^{232}Th$
&6.0e-06&3.8e-06&5.2e-06&1.8e-05&4.0e-05&4.5e-05&2.7e-05&7.8e-06&1.1e-06&6.9e-08&1.5e-04\\
\hline
&$^{238}U$
&9.4e-06&3.0e-06&6.1e-07&8.1e-08&6.8e-09&3.6e-10&1.2e-11&2.6e-13&3.4e-15&0.0&1.3e-05\\
\raisebox{1.5ex}{$6$}&$^{232}Th$
&2.1e-09&6.1e-10&7.5e-09&6.7e-08&4.0e-07&1.6e-06&4.2e-06&7.5e-06&9.1e-06&7.6e-06&3.0e-05\\
\hline
&$^{238}U$
 &0.0 &0.0 &0.0 &0.0 &0.0 &0.0 &0.0 &0.0 &0.0 &0.0 &0.0\\
\raisebox{1.5ex}{$>7$}&$^{232}Th$
&4.3e-06&1.6e-06&4.2e-07&7.2e-08&8.2e-09&6.1e-10&3.0e-11&9.7e-13&2.0e-14&0.0&6.4e-06\\
\hline
&$^{238}U$
&\multicolumn{10}{r}{}&1.6e-01\\
\raisebox{1.5ex}{Tot}&$^{232}Th$
 &\multicolumn{10}{r}{}&1.7e-01\\
\hline
\end{tabular}
\label{rateFe}
\end{sidewaystable}

\begin{sidewaystable}
\caption{Neutron production rate via $(\alpha, n)$ reaction due to natural radioactivity of samarium, specifically. }
\centering
\begin{tabular}{ l|ccccccccccc}\hline
Range$(MeV)$ &$0\sim .1$ &$.1\sim .2$
                                &$.2\sim .3$ &$.3\sim .4$ &$.4\sim .5$ &$.5\sim .6$
                                &$.6\sim .7$ &$.7\sim .8$ &$.8\sim .9$ &$.9\sim 1$  &Sum \\
\hline
 $E_{n}$  & \multicolumn{11}{c}{$\alpha + O \rightarrow $
                         Outgoing Neutron Flux $(ppm^{-1}g^{-1}year^{-1})$}  \\
\hline
0
 & /       &8.8e-14 &1.0e-13 &9.9e-14 &1.8e-12 &1.5e-09 &1.0e-07 &4.2e-07 &4.3e-07 &2.4e-07 &1.2e-06\\
1
 &4.1e-07 &5.2e-07 &3.0e-07 &7.3e-08 &6.9e-09 &3.9e-10 &3.8e-10 &7.8e-10 &1.5e-09 &2.6e-09 &1.3e-06\\
2
 &4.1e-09 &5.9e-09 &7.7e-09 &9.2e-09 &9.9e-09 &9.8e-09 &8.8e-09 &7.2e-09 &5.4e-09 &3.6e-09 &7.1e-08\\
3
 &2.3e-09 &1.3e-09 &6.5e-10 &3.0e-10 &1.3e-10 &5.0e-11 &1.7e-11 &5.6e-12 &1.6e-12 &4.3e-13 &4.7e-09\\
$>4$
 &1.0e-13 &2.3e-14 &4.5e-15 &0.0 &0.0 &0.0 &0.0 &0.0 &0.0 &0.0 &1.3e-13\\

\hline
Tot &\multicolumn{10}{r}{}&2.6e-06\\
\hline
\end{tabular}
\label{rateOSm}
\end{sidewaystable}

\begin{sidewaystable}
\caption{Neutron production rate via $(\alpha, n)$ reaction due to natural radioactivity of samarium, specifically. }
\centering
\begin{tabular}{ l|ccccccccccc}\hline
Range$(MeV)$ &$0\sim .1$ &$.1\sim .2$
                                &$.2\sim .3$ &$.3\sim .4$ &$.4\sim .5$ &$.5\sim .6$
                                &$.6\sim .7$ &$.7\sim .8$ &$.8\sim .9$ &$.9\sim 1$  &Sum \\
\hline
 $E_{n}$  & \multicolumn{11}{c}{$\alpha + Si \rightarrow $
                         Outgoing Neutron Flux $(ppm^{-1}g^{-1}year^{-1})$}  \\
\hline
0
 & /       &1.5e-13 &1.5e-13 &7.1e-09 &7.5e-09 &3.4e-10 &0.0 &0.0 &0.0 &0.0 &1.5e-08\\
$ >1 $& \multicolumn{10}{r}{}&0.0\\
\hline
Tot &\multicolumn{10}{r}{}&1.5e-08\\
\hline
\end{tabular}
\label{rateSiSm}
\end{sidewaystable}

\begin{sidewaystable}
\caption{Neutron production rate via $(\alpha, n)$ reaction due to natural radioactivity of samarium, specifically. }
\centering
\begin{tabular}{ l|ccccccccccc}\hline
Range$(MeV)$ &$0\sim .1$ &$.1\sim .2$
                                &$.2\sim .3$ &$.3\sim .4$ &$.4\sim .5$ &$.5\sim .6$
                                &$.6\sim .7$ &$.7\sim .8$ &$.8\sim .9$ &$.9\sim 1$  &Sum \\
\hline
 $E_{n}$  & \multicolumn{11}{c}{$\alpha + B \rightarrow $
                         Outgoing Neutron Flux $(ppm^{-1}g^{-1}year^{-1})$}  \\
\hline
0
& /       &1.5e-10 &1.9e-05 &1.9e-05 &1.8e-10 &1.2e-09 &1.0e-08 &7.9e-08 &5.0e-07 &2.6e-06 &4.1e-05\\
1
 &1.1e-05 &4.1e-05 &1.2e-04 &2.9e-04 &5.7e-04 &9.4e-04 &1.3e-03 &1.4e-03 &1.3e-03 &9.8e-04 &6.9e-03\\
2
 &6.1e-04 &3.2e-04 &1.5e-04 &6.6e-05 &3.7e-05 &2.9e-05 &2.7e-05 &2.6e-05 &2.4e-05 &2.1e-05 &1.3e-03\\
3
 &1.8e-05 &1.4e-05 &1.0e-05 &7.3e-06 &4.9e-06 &3.1e-06 &1.9e-06 &1.1e-06 &5.8e-07 &3.0e-07 &6.1e-05\\
4
 &1.4e-07 &6.7e-08 &2.9e-08 &1.2e-08 &4.7e-09 &1.8e-09 &6.2e-10 &2.1e-10 &6.5e-11 &2.0e-11 &2.6e-07\\
$>5$
 &5.6e-12 &1.5e-12 &3.8e-13 &9.2e-14 &2.1e-14 &4.6e-15 &9.5e-16 &0.0 &0.0 &0.0 &7.6e-12\\

\hline
Tot &\multicolumn{10}{r}{}&8.3e-03\\
\hline

\end{tabular}
\label{rateBSm}
\end{sidewaystable}

\newpage

\begin{figure}
\centering
\includegraphics[width=0.9\textwidth]{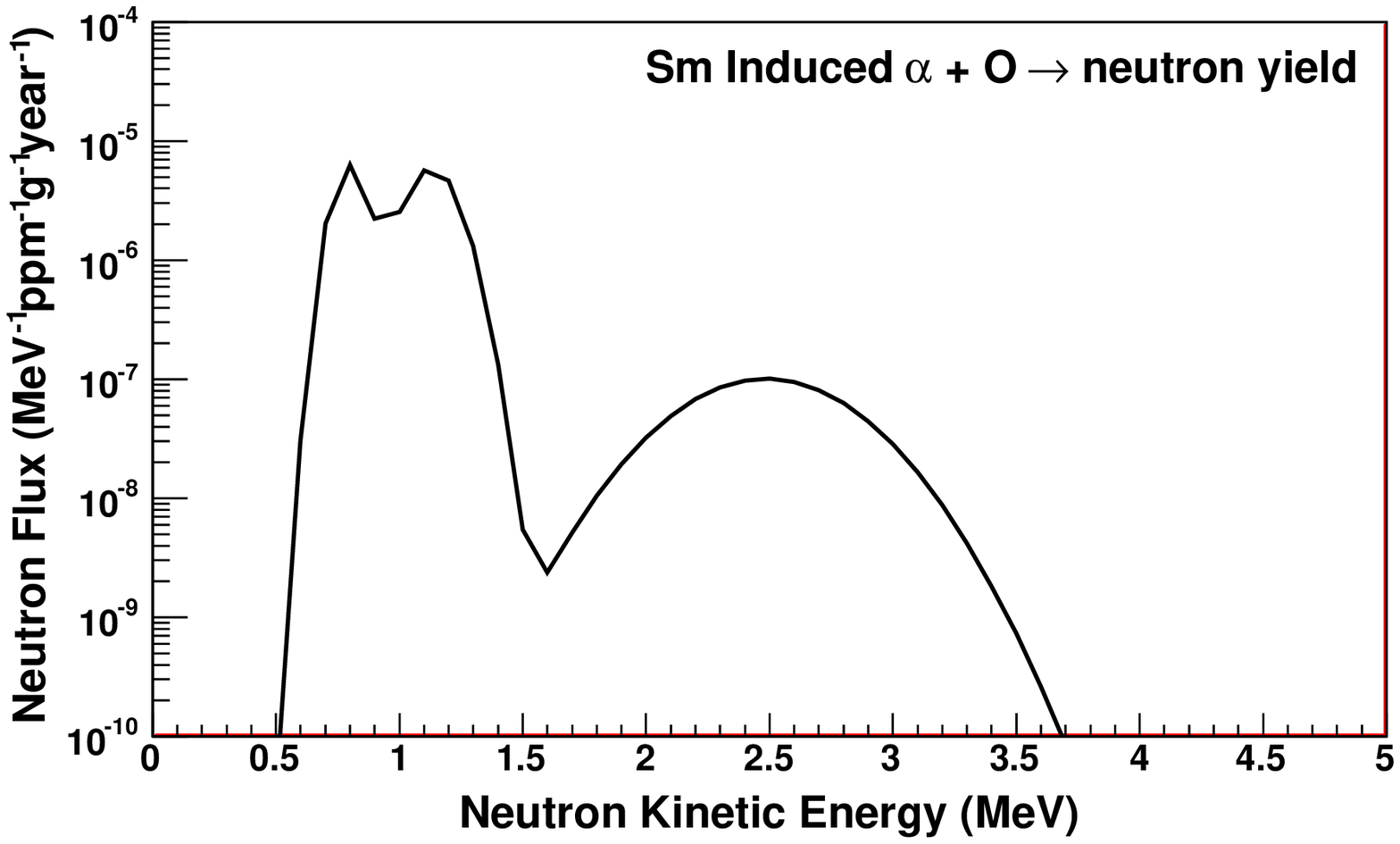}
\caption{ The differential neutron flux induced by $(\alpha, n)$ reaction in a thick target of oxygen. The $\alpha$-particles
  are induced by samarium decays.}\label{fig:OSm}
\end{figure}

\begin{figure}
\centering
\includegraphics[width=0.9\textwidth]{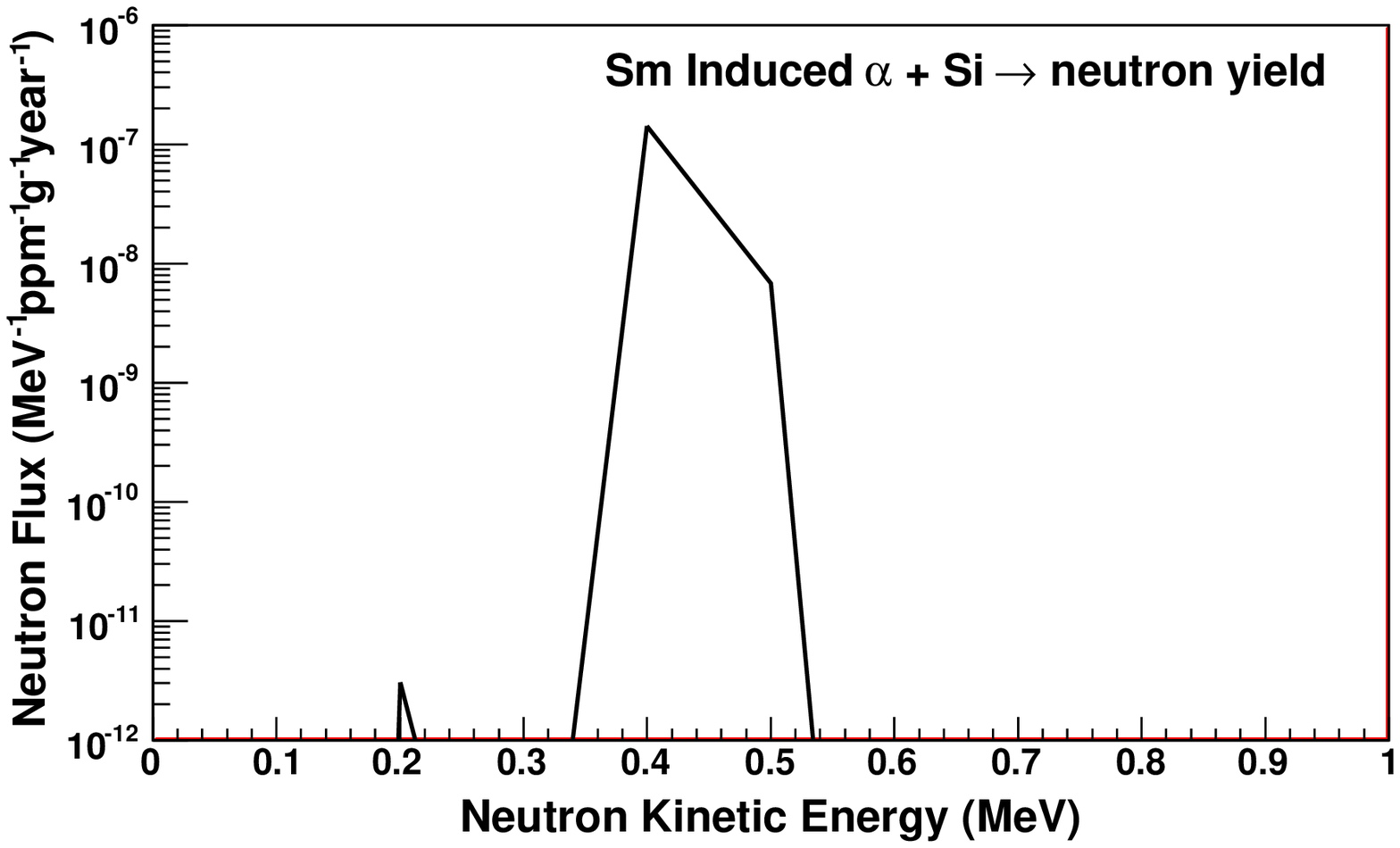}
\caption{The differential neutron flux induced by $(\alpha, n)$ reaction in a thick target of silicon. The $\alpha$-particles
  are induced by samarium decays.}\label{fig:SiSm}
\end{figure}

\begin{figure}
\centering
\includegraphics[width=0.9\textwidth]{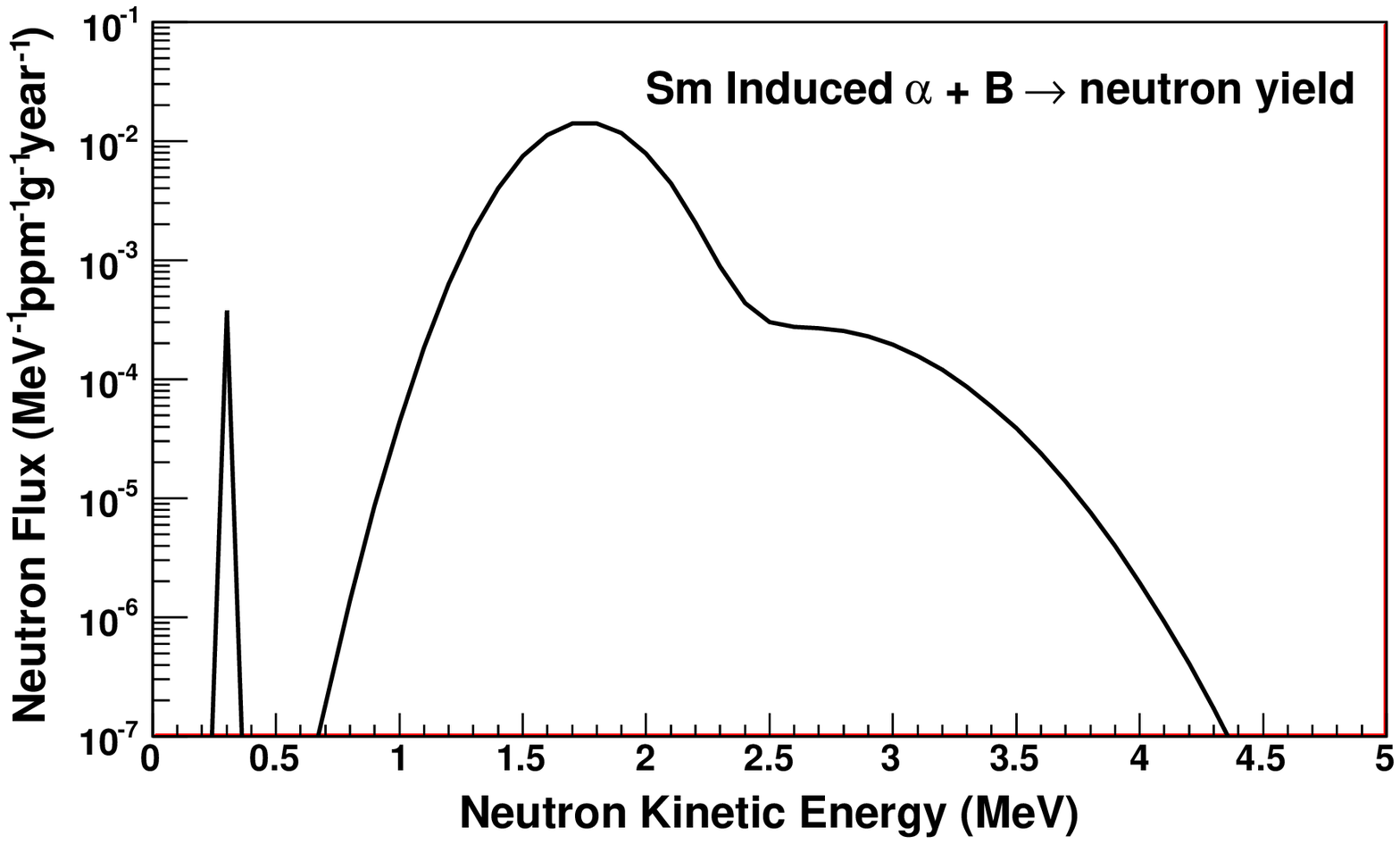}
\caption{The differential neutron flux induced by $(\alpha, n)$ reaction in a thick target of boron. The $\alpha$-particles
  are induced by samarium decays.}\label{fig:BSm}
\end{figure}
\section{Conclusions}
We have calculated the ($\alpha$,n) neutron yield and energy spectrum for many elements that are used to build low background experiments. Both neutron yields and 
energy spectra are important to the contribution in the energy region of interest. Therefore, it is critical to have information from both in the Monte Carlo simulation to predict 
the possible contributions in the energy region of interest from the ($\alpha$,n) neutrons. The neutron yield from many elements are compared to Heaton {\it et al.}~\cite{heat1990}. Good overall agreement is obtained. 
  
\section{Acknowledgment}
The authors wish to thank Yongchen Sun and Christina Keller at The University of
South Dakota for the invaluable support that made this work successful. 
This work was supported in part by the NSF grant 0758120, the Office of Research at The University 
of South Dakota, and by Laboratory Directed Research 
and Development at Los Alamos National Laboratory.


\begin{thebibliography}{99}
\bibitem{dns}D. N. Spergel {\it et al.}, Astrophys. J. Suppl. Ser. {\bf 148}, 175 (2003).
\bibitem{wfr}W. Freeman and M. Turner, Rev. Mod. Phys. {\bf 75}, 1433 (2003).
\bibitem{mwg}M. W. Goodman and E. Witten, Phys. Rev.  {\bf D 31}, 3059 (1985).
\bibitem{gjm}G. Jungman, M. Kamionkowski, and K. Griest, Phys. Rep. {\bf 267}, 195 (1996).
\bibitem{cdms1} D. S. Akerib {\it et al.} (CDMS Collaboration), Phys. Rev. Lett. {\bf 93}, 211301 (2004).
\bibitem{edel1} A. Benoit {\it et al.} (EDELWEISS Collaboration), Phys. Lett. {\bf B 545}, 43 (2002). 
\bibitem{xenon10} J. Angle {\it et al.} (XENON10 Collaboration), Phys. Rev. Lett. {\bf 100}, 021303 (2008).
\bibitem{argon} P. Benetti {\it et al.}, Astroparticle Physics, V {\bf 28}, 495 (2008).
\bibitem{dama1} R. Bernabei {\it et al.} (DAMA Collaboration),  Phys. Lett. {\bf B 480}, 23 (2000).
\bibitem{cresst1} G. Angloher {\it et al.}, Astropart. Phys. {\bf 18}, 43 (2002).
\bibitem{pica1}  M. Barnabe-Heider {\it et al.} (PICASSO), Phys. Lett. {\bf B 624}, 186 (2005).
\bibitem{naia} G. J. Alner {\it et al.} (UKDMC), Phys. Lett. {\bf B 616}, 17 (2005).
\bibitem{zep1}  V. A. Kudryavtsev (UKDMC), in the Fifth International Workshop on the Identification of Dark Matter, Edinburgh, Scotland, 2004.
\bibitem{bern1} R. Bernabei {\it et al.} (DAMA Collaboration), Riv. N. Cim. {\bf 26} (2003) 1-73.
\bibitem{bern2} R. Bernabei {\it et al.} (DAMA Collaboration), arXiv:0804.2741v1.
\bibitem{cdms2}Z. Ahmed et al., (CDMS Collaboration), pre-print, astro-ph/0802.3530
%\bibitem{mei1} D.-M. Mei, Z.-B. Yin, and A. Hime, In preparation.
\bibitem{wes} D. West and A. C. Sherwood, Ann. Nucl. Eng. {\bf 9}, 551 (1982).
\bibitem{bai} J. K. Bai and J. Gomez del Campo, Nucl. Sci. Eng. {\bf 71}, 18 (1978).
\bibitem{lis} H. Lisien and A. Paulsen, Atomkernenergie {\bf 30}, 1 (1970).
\bibitem{fei}Y. Feige, B. G. Oltman, and J. Kastner, J. Geophys. Res. {\bf 73}, 3135 (1968).
\bibitem{heat1990} R. Heaton {\it et al.}, Nucl. Geophys. V {\bf 4}, 499 (1990). 
\bibitem{heat} R. Heaton {\it et al.}, Nucl. Instrum. Methods Phys. Res. A {\bf 276}, 529 (1989). 
\bibitem{sha} S. Harissopulos {\it et al.}, Phys. Rev. C {\bf 72}, 062801 (2005).
\bibitem{cjh} G. J. H. Jacobs and H. Liskien, Ann. nucl. Energy, {\bf 10}, 541 (1983).
\bibitem{jhg} J. H. Gibbons and R. L. Macklin, Phys. Rev. {\bf 114}, 571 (1959).
\bibitem{wfi} W. Fitz, F. Kienle, R. Maschuw, and B. Zeitnitz, Phys. Rev. C {\bf 14}, 755 (1976).
\bibitem{lva} L. Van Der Zwan and K. W. Geiger, Nucl. Phys. {\bf A 246}, 93 (1975).
\bibitem{kev} Eric B. Norman,  Timothy E. Chupp, Kevin T. Lesko, Peter Schwalbach, and Patrick J. Grant, Nucl. Phys. A {\bf 390}, 561 (1982).
\bibitem{erb} Eric B. Norman,  Timothy E. Chupp, Kevin T. Lesko,  Patrick J. Grant, and Gene L. Woodruff, Phys. Rev. C {\bf 30}, 1339 (1984).
\bibitem{rou} R. K. Heaton, H. W. Lee, B. C. Robertson, E. B. Norman, K. T. Lesko, and B. Sur, NIM A 364 (1995), 317.
\bibitem{branch} D. S. Delion, A. Insolia, and R. J. Liotta, Nucl. Phys. (Supplement) A {\bf 654}, 673c (1999).  A. M. Sanchez and P. R. Montero, 
Nucl. Instrum. Methods Phys. Res. A {\bf 420}, 481 (1999).
\bibitem{talys}  A. J. Koning, S. Hilaire and M. C. Duijvestijn,
``TALYS: Comprehensive nuclear reaction modeling,'' Proceedings of the
International Conference on Nuclear Data for Science and Technology -
ND2004, AIP vol. 769, eds. R. C. Haight, M. B. Chadwick, T. Kawano,
and P. Talou, Sep. 26-Oct. 1, 2004, Sante Fe, USA, 2005, pp. 1154.
\bibitem{andr} Andreas Hermann and Doris Ehrt, Journal of Non-Crystalline Solids, V {\bf 354}, 916 (2008).
\bibitem{msio} M. S. Iovu {\it et al.}, Journal of Optoelectronics and Advanced Materials V {\bf 8}, 1341 (2006).
\bibitem{mei} D.-M. Mei {\it et  al.}, Astroparticle Physics 30, (2008) 12-17. 
\bibitem{astar} http://physics.nist.gov/PhysRefData/Star/Text/ASTAR.html.
\end{thebibliography}
\end{document}